\documentclass[fleqn,usenatbib,usedcolumn]{mnras}

\usepackage{newtxtext,newtxmath}

\usepackage[T1]{fontenc}
\usepackage{ae,aecompl}

\usepackage{graphicx}	
\usepackage{amsmath}	
\usepackage{amssymb}	
\usepackage{ulem}	
\usepackage{caption}
\usepackage[flushleft,referable]{threeparttablex}
\usepackage{footnote}
\usepackage[mathscr]{euscript}

\usepackage{floatrow}
\newcommand{\OIII}{[{\sc O iii}]\,$\lambda$}
\newcommand{\OII}{[{\sc O ii}]\,$\lambda$}
\newcommand{\logOH}{\log({\rm O/H})}
\newcommand{\MTO}{M_{\rm TO}}
\newcommand{\MZ}{$M_{\star}$--$Z$}
\newcommand{\MACT}{$\mathcal{MACT}$}
\newcommand{\Msun}{$M_{\odot}$}
\newcommand{\logM}{$\log(M_{\star}/M_{\odot})$}
\newcommand{\zevo}{$(1+z)^{-1.45^{+0.61}_{-0.76}}$}
\newcommand{\z}{$z\approx0.8$}

\defcitealias{AM13}{AM13}
\defcitealias{Zahid13}{Z13}
\defcitealias{BC03}{BC03}
\defcitealias{Maraston05}{M05}
\defcitealias{Maraston11}{M11}
\defcitealias{Kobulnicky04}{KK04}
\defcitealias{Comparat17}{C17}

\title[Stellar Populations of $z\approx0.8$ Metal-Poor Galaxies]{The Stellar Population of Metal-Poor Galaxies at \z\ and the Evolution of the Mass--Metallicity Relation}

\author[A. Weldon, C. Ly, \& M. Cooper]{
Andrew Weldon,$^{1}$\thanks{E-mail: weldon119@email.arizona.edu}
Chun Ly,$^{1}$
and Michael Cooper$^{2}$\\
$^{1}$Steward Observatory, University of Arizona, 933 N Cherry Avenue, Tucson, AZ 85721, USA\\
$^{2}$Center for Cosmology, Department of Physics \& Astronomy, 4129 Reines Hall, University of California, Irvine, CA 92697, USA}

\date{Accepted 2019 October 29. Received 2019 October 28; in original form 2019 April 11}

\pubyear{2019}

\begin{document}
\label{firstpage}
\pagerange{\pageref{firstpage}--\pageref{lastpage}}
\maketitle

\begin{abstract}
  We present results from deep \textit{Spitzer}/Infrared Array Camera (IRAC) observations of 28 metal-poor, strongly star-forming galaxies selected from the DEEP2 Galaxy Survey. By modelling infrared and optical photometry, we derive stellar masses and other stellar properties.
  We determine that these metal-poor galaxies have low stellar masses, $M_{\star}$ $\approx10^{8.1}$--10$^{9.5}$ \Msun. Combined with the Balmer-derived star formation rates (SFRs), these galaxies have average inverse SFR/$M_{\star}$ of $\approx$100 Myr.
  The evolution of stellar mass--gas metallicity relation to \z\ is measured by combining the modelled masses with previously obtained spectroscopic measurements of metallicity from \OIII 4363 detections. Here, we include measurements for 79 galaxies from the Metal Abundances across Cosmic Time Survey. Our mass--metallicity relation is lower at a given stellar mass than at $z=0.1$ by 0.27 dex. This demonstrates a strong evolution in the mass--metallicity relation, \zevo.
  We find that the shape of the \z\ mass-metallicity relation, a steep rise in metallicity at low stellar masses, transitioning to a plateau at higher masses, is consistent with $z\sim0.1$ studies. We also compare the evolution in metallicity between \z\ and $z\sim0.1$ against recent strong-line diagnostic studies at intermediate redshifts and find good agreement. Specifically, we find that lower mass galaxies ($4\times10^8$ \Msun) built up their metal content 1.6 times more rapidly than higher mass galaxies ($10^{10}$ \Msun). Finally, we examine whether the mass--metallicity relation has a secondary dependence on SFR, and statistically concluded that there is no strong secondary dependence for \z\ low-mass galaxies.
\end{abstract}




\section{Introduction}
One key observable for understanding galaxy evolution is the chemical content of the interstellar gas and how it evolves. Specifically, comparisons of the gas-phase heavy-element abundances (i.e., ``metallicity'') against the stellar mass enable us to understand how baryons behave in galaxies \citep{Tremonti04}. The chemical enrichment of galaxies is first driven by star formation. As massive stars age, they produce heavy elements in their cores which are released via supernova winds into the interstellar medium when these stars reach their end stage. This chemical enrichment process is regulated by the cosmic accretion of more chemically pristine gas \citep[e.g.,][]{Dave11, Lilly13}.

The most reliable chemical abundance measurements are derived from the electron temperature ($T_e$) method \citep{Aller84,Izotov06}. This method uses the emission-line flux ratio of \OIII 4363 to \OIII $\lambda$4959,5007.
However, reliable measurements of \OIII 4363 are difficult to obtain in metal-rich and even harder to detect in high-$z$ galaxies. Currently, there are 172 intermediate-redshift galaxies with robust (S/N $\geq3$) $T_e$-based metallicities \citep[e.g.,][]{Hoyos05, Hu09, Atek11, Amorin14, Amorin15, Jones15, Ly2014, Ly2015b, Ly2016a}.\\
\indent Due to the difficulty in obtaining temperature-based metallicities, previous studies have been primarily limited to using strong nebular emission lines \citep[e.g.,][]{Pagel79, Pettini04} to measure the evolution of the mass--metallicity (\MZ) relation \citep[e.g.,][and references therein]{Erb06, Liu08, Maiolino08, Lamareille09, Thuan10, Hunt12, Nakajima12, Wuyts12, Yabe12, Yabe14, Yabe15, Yates12, Belli13, Guaita13, Henry13, Cullen14,  Maier14, Salim14, Troncoso14, Whitaker14, Zahid14, Reyes15, Sanders15}. A number of mass--metallicity studies have used the $T_e$ method, but mostly in the local universe with stacked spectra \citep[e.g.,][hereafter AM13]{AM13}. In addition, robust stellar mass measurements at $z\gtrsim1$ require deep infrared observations.\\
\indent To address the lack of reliable $T_e$-based metallicity and stellar mass measurements at \z, we present new infrared measurements from the {\it Spitzer Space Telescope} \citep[hereafter \textit{Spitzer}]{Werner04} for 28 \z\ \OIII 4363-detected galaxies identified by \cite{Ly2015b} from the DEEP2 survey \citep{Newman13}.

The primary objectives of this study are: (1) measure the stellar population (i.e., stellar mass, stellar age, SFR) of these metal-poor galaxies; (2) determine the relationship between stellar mass and gas metallicity at \z\ for low-mass galaxies; and (3) explore the possible relationship between SFR, stellar mass, and gas metallicity. The oxygen abundances for these galaxies have been determined using the $T_e$ method by \cite{Ly2015b}. Currently, the stellar masses of these galaxies are poorly constrained with existing optical data, 1900--5000 \AA\ rest-frame. \textit{Spitzer}/IRAC \citep{Fazio04} observations, which sample red-ward of 1.6 $\micron$, are necessary for reliable stellar mass constraints.
    
This paper is outlined as follows. In Section~\ref{sec:2}, we discuss the DEEP2 sample, the \textit{Spitzer} data for this study, and the Metal Abundance across Cosmic Time (\MACT) survey \citep{Ly2016a}. The latter is used to extend our analyses toward lower stellar masses relative to the DEEP2 sample. In Section~\ref{sec:3}, we describe our approach for estimating stellar masses and other stellar population properties from spectral energy distribution (SED) modelling. In Section~\ref{sec:4}, we present our main results on the stellar population for these low-metallicity galaxies, our \MZ\ relation, the evolution of the \MZ\ relation, its dependence on SFR, and the selection function of our study. We summarize our results in Section~\ref{sec:5}.

Throughout this paper, we adopt a flat cosmology with $\Omega_{\Lambda}$ = 0.728, $\Omega_{\rm M}$ = 0.272, and $H_{0}$ = 70.4 km s$^{-1}$ Mpc$^{-1}$ \citep[WMAP7;][]{Komatsu11}.\footnote{WMAP7 is the default cosmology used by our modelling software, see Section \ref{sec:modelling}.} We note that adopting $\Omega_{\Lambda}$ = 0.7, $\Omega_{\rm M}$ = 0.3, and $H_{0}$ = 70 km s$^{-1}$ Mpc$^{-1}$ would decrease stellar masses by 0.007 dex at $z=0.8$. Magnitudes are reported on the AB system \citep{Oke1974}.

\section{The Samples}
\label{sec:2}

\subsection{DEEP2 Survey}
In this study, we use the sample of 28 \z\ galaxies identified by \cite{Ly2015b} to have spectroscopic detections of \OIII 4363. Here, we briefly summarize the sample and refer readers to \cite{Ly2015b} for more details. From the fourth data release\footnote{\url{http://deep.ps.uci.edu/dr4/}} of the DEEP2 Galaxy Redshift Survey ($\approx$3 deg$^2$ surveyed area), 37,396 sources were selected with reliable spectroscopic redshifts. Of those, only sources with rest-frame spectral coverage of 3720--5010 \AA\ were considered, limiting the sample to 4,140 galaxies.
The emission lines of these galaxies were then fitted with Gaussian profiles, following the approach of \cite{Ly2014}. Using emission-line fluxes, galaxies were selected with \OIII 4363 and \OIII 5007 detections at S/N $\geq$ 3 and S/N $\geq$ 100, respectively. Of the remaining galaxies, 26 were removed primarily due to OH night skyline contamination. This reduced the sample to 28 \OIII 4363-detected galaxies at $z=0.710$--0.856, with a median (average) redshift of 0.774 (0.773).

\subsubsection{\textit{Spitzer}/IRAC Data}
\begin{figure*}
  \centering
  \includegraphics[width=7.25in,keepaspectratio]{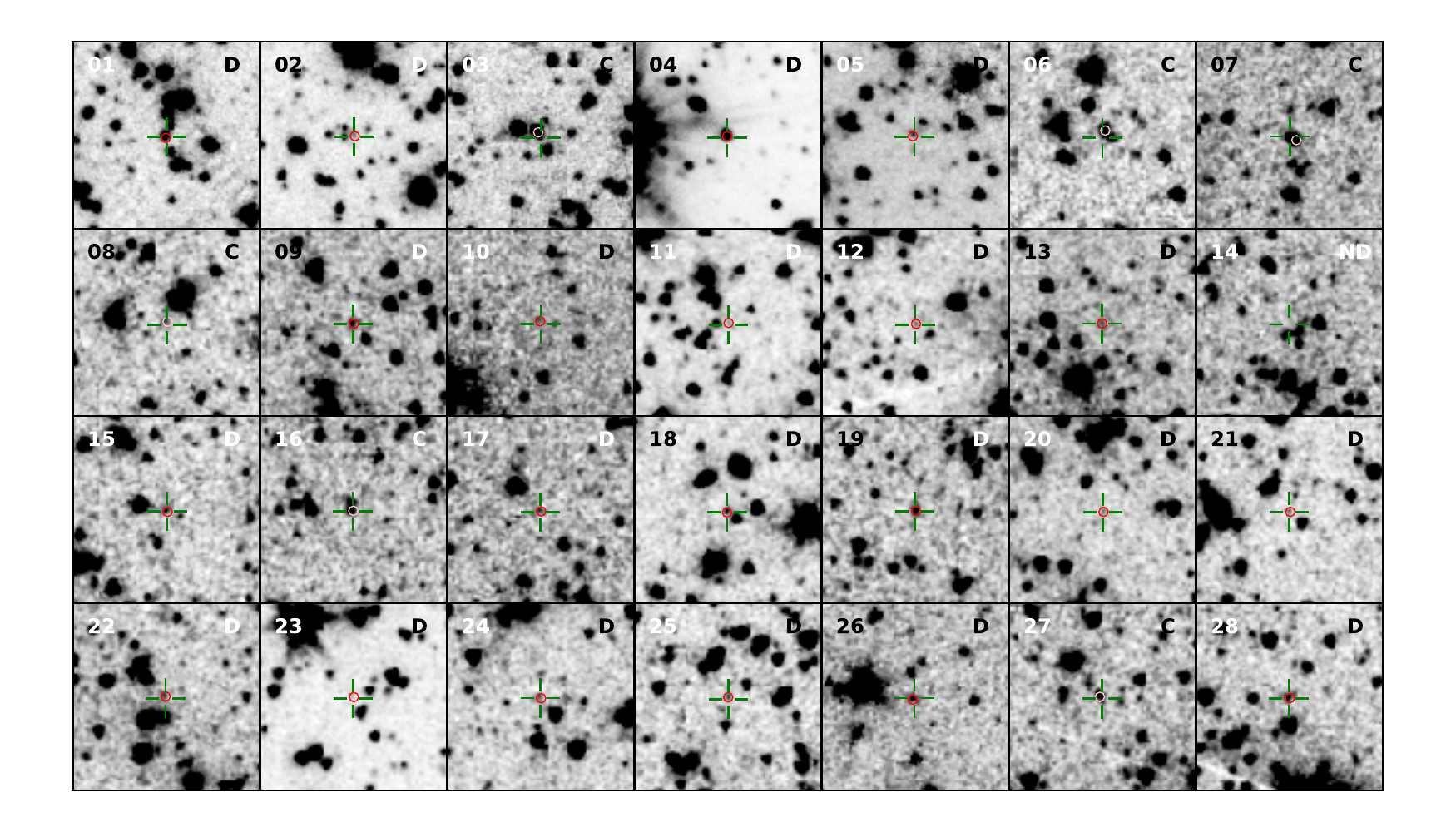}
  \includegraphics[width=7.25in,keepaspectratio]{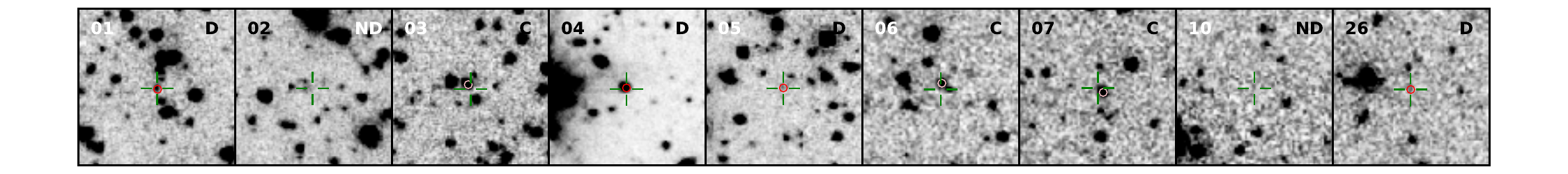}
  \vspace{-0.5cm}
  \caption{\textit{Spitzer}/IRAC 3.6 $\mu$m (4.5 $\mu$m) imaging for 28 (9) metal-poor galaxies illustrated on top (bottom) panel. Of the galaxies with 3.6 $\mu$m (4.5 $\mu$m) data, 21 (four) were detected (``D''), one (two) was not detected (``ND''), and six (three) suffer from confusion/blending with nearby sources (``C''). The targeted galaxies are at the centre of the images. Green crosshairs mark the optical position of the galaxies. These positions were improved by \citet{Matthews13} using SDSS and CFHT data. Red circles (radius of 1\farcs5) mark the IRAC position, and pink circles (radius of 1\farcs5) mark the closest object found by MOPEX to the optical position for blended cases. Each image is 40\arcsec\ on a side, which corresponds to an angular distance of 300 kpc at $z=0.8$, and oriented such that North is up and East is left.}
  \label{fig:stamps}
\end{figure*}

Of the 28 DEEP2 galaxies, 19 galaxies had no previous \textit{Spitzer} data, or the existing data were too shallow for stellar mass estimates. In order to derive robust stellar masses, we observe these 19 galaxies at 3.6 $\mu$m with the \textit{Spitzer}/IRAC \citep[Prop. No. 12104;][]{Ly2015a}. For DEEP2 galaxies \#01--07, 10 and 26, see Table \ref{tab:DEEP2}, existing \textit{Spitzer} 3.6 $\mu$m and 4.5 $\mu$m data were obtained through the \textit{Spitzer} Heritage Archive.\footnote{\url{http://sha.ipac.caltech.edu/applications/Spitzer/SHA/}}
We illustrate the \textit{Spitzer}/IRAC 3.6 $\mu$m and 4.5 $\mu$m data in Figure \ref{fig:stamps}.
The total integration times are listed in Table \ref{tab:DEEP2}. The median (mean) total integration times were 52.5 (52.54) minutes for 3.6 $\mu$m and 47 (75.7) minutes for 4.5 $\mu$m.

\newcommand{\FA}{\tnotex{tn:1}}
\newcommand{\FB}{\tnotex{tn:2}}
\newcommand{\FC}{\tnotex{tn:3}}
\newcommand{\FD}{\tnotex{tn:4}}
\newcommand{\FE}{\tnotex{tn:5}}
\newcommand{\PA}{\phantom{1}}
\newcommand{\PB}{\phantom{11}}
\begin{table*}
  \centering
  \caption{Summary of \textit{Spitzer}/IRAC Observations and Photometry}
  \label{tab:DEEP2}
  \begin{threeparttable}
    \begin{tabular}{ccccccccc}
      \hline\hline
      ID & R.A.        & Declination & Obs. Date  & Int. Time 			& $F$([3.6])\FA & $F$([4.5])\FA 	& IRAC Det. 		& Aper. Corr. \\
         & hh:mm:ss.ss  & dd:mm:ss.ss  &        & (min)     			& ($\mu$Jy)     & ($\mu$Jy)   		& 					&	 \\
      (1)& (2)         & (3)         & (4)        & (5)       			& (6)           & (7)   			& 	(8)				&	(9)\\\hline
      01 & 14:18:31.26 & 52:49:42.49 & \ldots\FB  & 227/230\FC    	& 4.30$\pm$0.14 & 2.86$\pm$0.18		&	Yes/Yes	  		& 	1.55/1.53\\
      02 & 14:21:21.53 & 53:01:07.84 & \ldots\FB  & 113/127\FC    	& 2.47$\pm$0.22 & $<$0.64\FD		&	Yes/No   		& 	1.55/1.50\\
      03 & 14:21:25.48 & 53:09:48.09 & \ldots\FB  & 143/136\FC    	& \ldots        & \ldots			&	Conf./Conf. 	& 	1.55/1.62\\
      04 & 14:22:03.72 & 53:25:47.77 & \ldots\FB  & 103/125\FC    	& 35.04$\pm$0.24& 27.80$\pm$0.27 	&	Yes/Yes   		& 	1.55/1.64\\
      05 & 14:21:45.40 & 53:23:52.71 & \ldots\FB  & \PA52/\PA47\FC  & 3.49$\pm$0.18 & 2.58$\pm$0.20		&	Yes/Yes   		& 	1.57/1.66\\
      06 & 16:47:26.17 & 34:45:11.89 & \ldots\FB  & \PB 5/\PB 3\FC  & \ldots        & \ldots			&	Conf./Conf. 	& 	1.64/1.65\\
      07 & 16:46:35.40 & 34:50:27.82 & \ldots\FB  & \PB 3/\PB 3\FC  & \ldots        & \ldots			&	Conf./Conf. 	& 	1.67/1.65\\
      08 & 16:47:26.47 & 34:54:09.55 & 2016-05-26 & \PA25/\ldots\FE     & \ldots        & \ldots			&	Conf./\ldots\FE & 	1.62/\ldots\FE\\
      09 & 16:49:51.35 & 34:45:18.05 & 2016-05-26 & \PA29/\ldots\FE     & 3.60$\pm$0.30 & \ldots			&	Yes/\ldots\FE   & 	1.54/\ldots\FE\\
      10 & 16:51:31.45 & 34:53:15.83 & \ldots\FB  & \PB 4/\PB 4\FC  & 2.73$\pm$0.36 & $<$1.87\FD		&	Yes/No   		& 	1.63/1.70\\
      11 & 16:50:55.32 & 34:53:29.66 & 2016-05-26 & \PA56/\ldots\FE     & 1.36$\pm$0.24 & \ldots			&	Yes/\ldots\FE   & 	1.47/\ldots\FE\\
      12 & 16:53:03.46 & 34:58:48.71 & 2016-05-26 & \PA56/\ldots\FE     & 1.30$\pm$0.40 & \ldots			&	Yes/\ldots\FE   & 	1.51/\ldots\FE\\
      13 & 16:51:24.04 & 35:01:38.54 & 2016-05-26 & \PA26/\ldots\FE   	& 2.02$\pm$0.30 & \ldots			&	Yes/\ldots\FE   & 	1.57/\ldots\FE\\
      14 & 16:51:20.32 & 35:02:32.39 & 2016-05-26 & \PA26/\ldots\FE    	& $<$0.92\FD    & \ldots			&	No/\ldots\FE    & 	1.58/\ldots\FE\\
      15 & 23:27:20.38 & 00:05:54.40 & 2016-03-07 & \PA56/\ldots\FE     & 1.77$\pm$0.30 & \ldots			&	Yes/\ldots\FE   & 	1.67/\ldots\FE\\
      16 & 23:27:43.14 & 00:12:42.45 & 2016-03-07 & \PA14/\ldots\FE     & \ldots        & \ldots			&	Conf./\ldots\FE & 	1.74/\ldots\FE\\
      17 & 23:27:29.87 & 00:14:19.94 & 2016-03-07 & \PA14/\ldots\FE     & 2.03$\pm$0.40 & \ldots			&	Yes/\ldots\FE   & 	1.67/\ldots\FE\\
      18 & 23:27:07.50 & 00:17:41.16 & 2016-03-07 & \PA30/\ldots\FE     & 4.39$\pm$0.50 & \ldots			&	Yes/\ldots\FE   & 	1.72/\ldots\FE\\
      19 & 23:26:55.43 & 00:17:52.68 & 2016-09-14 & \PA56/\ldots\FE     & 3.63$\pm$0.36 & \ldots			&	Yes/\ldots\FE   & 	1.77/\ldots\FE\\
      20 & 23:30:57.95 & 00:03:37.92 & 2016-09-13 & \PA56/\ldots\FE     & 1.15$\pm$0.39 & \ldots			&	Yes/\ldots\FE   & 	1.79/\ldots\FE\\
      21 & 23:31:50.74 & 00:09:38.91 & 2016-09-14 & \PA56/\ldots\FE     & 1.01$\pm$0.37 & \ldots			&	Yes/\ldots\FE   & 	1.62/\ldots\FE\\
      22 & 02:27:48.86 & 00:24:39.57 & 2016-10-27 & \PA56/\ldots\FE     & 1.76$\pm$0.31 & \ldots			&	Yes/\ldots\FE   & 	1.62/\ldots\FE\\
      23 & 02:27:05.71 & 00:25:21.54 & 2016-10-27 & \PA45/\ldots\FE     & 1.81$\pm$0.27 & \ldots			&	Yes/\ldots\FE   & 	1.75/\ldots\FE\\
      24 & 02:27:30.46 & 00:31:06.04 & 2016-10-27 & \PA51/\ldots\FE     & 1.81$\pm$0.37 & \ldots			&	Yes/\ldots\FE   & 	1.70/\ldots\FE\\
      25 & 02:26:03.72 & 00:36:21.98 & 2016-10-27 & \PA56/\ldots\FE     & 1.77$\pm$0.29 & \ldots			&	Yes/\ldots\FE   & 	1.63/\ldots\FE\\
      26 & 02:26:21.49 & 00:48:06.64 & \ldots\FB  & \PB 7/\PB 6\FC  & 5.11$\pm$0.28 & 4.33$\pm$0.57		&	Yes/Yes   		& 	1.49/1.67\\
      27 & 02:29:33.68 & 00:26:07.91 & 2016-10-27 & \PA53/\ldots\FE     & \ldots        & \ldots			&	Conf./\ldots\FE & 	1.62/\ldots\FE\\
      28 & 02:29:02.03 & 00:30:07.83 & 2016-10-27 & \PA53/\ldots\FE     & 3.15$\pm$0.30 & \ldots			&	Yes/\ldots\FE   & 	1.62/\ldots\FE\\
      \hline
    \end{tabular}
    \begin{tablenotes}
      \item \leavevmode\kern-\scriptspace\kern-\labelsep (1): DEEP2 \OIII 4363 galaxy ID. (2): Right ascension. (3): Declination. (4): Observation date for new \textit{Spitzer}/IRAC data. (5): Total on-source integration for 3.6 $\micron$ and 4.5 $\micron$ data, respectively. (6): \textit{Spitzer}/IRAC 3.6 $\micron$ flux. (7): \textit{Spitzer}/IRAC 4.5 $\micron$ flux. (8): Flags indicating whether galaxy was detected with \textit{Spitzer} or confused (``Conf.'') for 3.6 $\micron$ and 4.5 $\micron$, respectively. (9): Aperture corrections for 3.6 $\micron$ and 4.5 $\micron$, respectively.
      \item[a] \label{tn:1} Fluxes and uncertainties include aperture corrections given in Column (9).
      \item[b] \label{tn:2} These galaxies were observed multiple times over many years.
      \item[c] \label{tn:3} Integration times listed are the average exposure time from the \textit{Spitzer} Heritage Archive images.
      \item[d] \label{tn:4} 3$\sigma$ limits adopted.
      \item[e] \label{tn:5} IRAC 4.5 $\mu$m observations are not available.
    \end{tablenotes}
  \end{threeparttable}
\end{table*}

\subsection{\MACT\ Survey}
To increase the sample size and extend toward lower stellar masses ($M_{\star} \lesssim 3 \times10^9$ \Msun; DEEP2 is a magnitude-limited survey, $R < 24.1$), we include the \OIII 4363-based samples from the \MACT\ survey in our mass--metallicity relation analysis. Here, we briefly summarize the \MACT\ sample and refer readers to \cite{Ly2016a} for more details.

The survey targeted $\approx$1,900 emission-line galaxies of the Subaru Deep Field \citep[$\approx$0.25 deg$^2$ surveyed area;][]{Kashikawa04} that had excess flux in the narrow-band and/or intermediate-band filters \citep{Ly07}. Using Keck and MMT, deep rest-frame optical spectra were obtained. The \MACT\ survey consists of two sub-samples, those with detections of the \OIII 4363 line at S/N $\geq$ 3 ($N=29$ at $z\geq0.5$) and those with upper limits on \OIII 4363 ($N=50$; $z\geq0.5$). For the latter, \cite{Ly2016a} required that \OIII 5007 was detected at S/N $\geq$ 100 and \OIII 4363 at S/N $<$ 3. We refer to these two sub-samples collectively as the \MACT\ sample.

The stellar masses for the \MACT\ galaxies were determined with the Fitting and Assessment of Stellar Templates \citep[FAST;][]{Kriek09} code using the following photometric data: (1) FUV and NUV imaging from \textit{GALEX} \citep{Martin05}; (2) $U$-band imaging from Kitt Peak National Observatory (KPNO) Mayall telescope using MOSAIC \citep{Sawyer10}; (3) $BVR_{C}i'z'z_bz_r$ imaging from the Subaru telescope with Suprime-Cam \citep{Miyazaki02}; (4) $H$-band imaging from KPNO Mayall telescope using NEWFIRM \citep{Probst08}; and (5) $J$- and $K$-band imaging from UKIRT using WFCAM \citep{Casali07}.
Further discussion of the photometric data for \MACT\ is provided in \cite{Ly2011} and \cite{Ly2016a}. Together, the DEEP2 and \MACT\ galaxies are at $z = 0.578$ to $z = 0.955$ with a median (mean) redshift of 0.789 (0.763).

\section{Derived Measurements}
\label{sec:3}

\subsection{Infrared Photometry}
Using the MOsaicker and Point source EXtractor \citep[MOPEX, vers. 18.5.6;][]{MOPEX}, designed specifically to analyse \textit{Spitzer} data, we performed aperture photometry to obtain the stellar continuum emission from the DEEP2 \OIII 4363-detected galaxies. We executed MOPEX mostly with default parameters with the following differences: not executing the \texttt{fit\_radius} module, using a background-subtracted image for the point source probability (psp), adopting one aperture with a radius of 1\farcs8, and the values listed in Table \ref{tab:MOPEX} for the signal-to-noise ratio (\texttt{SNR}) in the \texttt{select\_conditions} constraint.\footnote{\texttt{SNR} is the threshold above which the targeted galaxy is detected by MOPEX.} We then applied aperture corrections to obtain total photometric fluxes and calculated the background rms flux to estimate the uncertainty in the \textit{Spitzer} images.

\newcommand{\FAc}{\tnotex{tn:1}}
\begin{table*}
  \centering
  \caption{Adopted SNR Threshold Values for Aperture Photometry with MOPEX}
  \label{tab:MOPEX}
  \begin{threeparttable}
    \begin{tabular}{cc}
      \hline\hline
      \texttt{SNR} & ID \\
      (1) & (2)\\
      \hline
      3.0 & 04, 05, 08, 09, 10, 11, 12, 15, 16, 19, 21, 22, 23, 24, 25, 26, 27, 28\\
      2.8 & 02, 03, 06, 07, 14, 18\\
      2.5 & 13, 17, 20\\
      2.0 & 01\\
      \hline
    \end{tabular}
    \begin{tablenotes}
    \item \leavevmode\kern-\scriptspace\kern-\labelsep (1):
      Signal-to-noise ratio threshold used to identify sources by MOPEX (set in the \texttt{select\_conditions} constraint). 
      (2): DEEP2 \OIII 4363 galaxy ID. An \texttt{SNR} of 2.5 was used for all {\it Spitzer} 4.5$\mu$m images.
    \end{tablenotes}
  \end{threeparttable}
\end{table*}

\subsubsection{Total \textit{Spitzer} Fluxes} 
The aperture corrections for the \textit{Spitzer}/IRAC 3.6 $\mu$m and 4.5 $\mu$m photometry were determined as follows:\\
First, we identified bright stars in each image (within a radius of 6\arcmin\ from the \OIII 4363-detected galaxy) using the coordinates of stars from the SDSS and cross-matched them against the \textit{Spitzer} MOPEX catalogues. Stars were then rejected if they were (1) within 12\arcsec\ of the edge of the image, (2) within 7\arcsec\ of another object identified by MOPEX, or (3) if the MOPEX flux was below 400 $\mu$Jy in a circular aperture (1\farcs8, radius). These restrictions were used to limit our calculations to bright isolated stars. Fluxes were then calculated using a circular aperture ranging from a radius of 1 pixel (0\farcs6) to 9 pixels (5\farcs4). The latter was chosen to include as much starlight as possible while excluding the light of other sources. The aperture corrections were then derived from the ratio of the 9- to 3-pixel fluxes. Outliers contaminated by nearby sources were removed iteratively using a sigma-clipping approach ($\sigma=2.5$). The aperture correction, given in Table \ref{tab:DEEP2}, is the mean from sigma clipping and has been applied to the IRAC fluxes in Table \ref{tab:DEEP2}. If there were four or fewer stars within the search area after the restrictions, the aperture correction was set to the average correction (1.62 for both bands) of the remaining images. The 3.6 (4.5) $\mu$m aperture corrections range from 1.47 to 1.79 (1.50 to 1.68).

\subsubsection{Measurements Uncertainties}

To determine the uncertainty in the IRAC 3.6 $\mu$m and 4.5 $\mu$m detections, we computed the background rms flux in the \textit{Spitzer} images. This was done by measuring the flux within randomly placed apertures throughout an image that has been masked for sources and then fitting the distribution of fluxes to obtain the Gaussian width. The Python package \texttt{photutils.detect} was used to construct a segmentation map to identify and mask pixels containing source emission. Here we required four contiguous pixels above a threshold value of 1.5$\sigma$. We then used \texttt{photutils.aperture\_photometry} to measure the flux within a \texttt{photutils.CircularAperture} of three pixels (1\farcs8, radius) for 12000 random positions across the masked image (typically $\sim$7000 positions were not affected by emission from sources). The distribution of background fluxes was then constructed and mirroring of the lower half relative to the median or peak (whichever was lower) was performed.\footnote{Mirroring limited the effect of under-masking as we found that the segmentation maps were not perfect.} For 3.6 $\mu$m data, the 3$\sigma$ flux limits range between 0.42 and 1.49 $\mu$Jy, with a median and mean of 0.91 and 0.93 $\mu$Jy, respectively. For 4.5 $\mu$m data, 3$\sigma$ flux limits range between 0.51 and 1.87 $\mu$Jy with a median (mean) of 0.72 (1.01) $\mu$Jy. These limits are provided in Table~\ref{tab:DEEP2}.

Additionally, we visually inspected the \textit{Spitzer} images to determine if the galaxy was detected and not blended with neighbouring sources. If the galaxy was not detected, we use the 3$\sigma$ background flux limit around the galaxy. If the galaxy was contaminated, we do not report a flux. Of the 28 galaxies, 21 were detected at 3.6 $\mu$m, one was not detected, and six suffered contamination from nearby sources. Among the nine galaxies with 4.5 $\mu$m data, four were detected, two were not detected, and three were confused with nearby sources.\\

\subsection{Stellar Masses}
\label{sec:modelling}
To estimate the stellar masses of DEEP2 \OIII 4363-detected galaxies, we combine \textit{Spitzer} measurements with optical and near-infrared data, to model the SED of each galaxy. We describe below the photometric data (Section \ref{sec:phot}), the adopted SED models (Section \ref{sec:models}), assumptions in our SED fitting (Section \ref{sec:model_reasons}), and stellar mass comparisons (Section \ref{sec:mass_comp}).

\subsubsection{Optical and Near-Infrared Data}
\label{sec:phot}
The optical and near-infrared data, provided in Table \ref{tab:phot}, include Canada-France-Hawaii Telescope (CFHT) $BRI$ photometry from \cite{Coil04}, $ugriz$ from \cite{Matthews13},\footnote{For DEEP2 \#01, 02, 03 and 05, $ugriz$ photometry are from the CFHT Legacy Survey \citep{Gwyn12}; all others are from SDSS.} $JHK_s$ for DEEP2 \#01 from the NEWFIRM Medium-Band Survey \citep{Whitaker11}, and $K_s$ photometry for DEEP2 \#04 from \cite{Bundy06}.

These photometric data have been corrected for nebular emission lines from optical spectroscopy using the technique described in \cite{Ly2015b} and \cite{Ly2016a}. For our sample, this correction can be significant due to the strong H$\beta$+[{\sc O iii}] emission such that it can over correct redder optical band(s), such as $z$-band.  To avoid this overcorrection, we perform the SED modelling excluding the affected band(s).

\subsubsection{SED Modelling}
\label{sec:models}
\textbf{Models.} The stellar synthesis models were created in Python using Code Investigating GALaxy Emission, \citep[CIGALE, vers. 0.9.0;][]{CIGALE}, an open-source software that has been well tested.\footnote{There is a growing trend in astronomy toward using open-source software, which helps enable scientific reproducibility \citep[e.g.,][]{Astropy}.} The models assumed a \citet[hereafter BC03]{BC03} simple stellar population (SSP), \cite{Chabrier03} initial mass function (IMF), and a ``delayed $\tau$'' star formation history:
\begin{equation}
	{\rm SFR}(t) \propto \frac{t}{\rm \tau_{main}^{2}}\exp{\left(\frac{-t}{\rm \tau_{main}}\right)},
\end{equation}
where ${\rm \tau_{main}}$ is the $e$-folding time of the main stellar population. Here, the SFR rises steadily, peaks at $t = {\rm \tau_{main}}$, and then declines exponentially.

In addition, we assumed a stellar metallicity of 0.02, a power-law dust attenuation with a slope of --0.7 \citep{Charlot00}, and an old-to-young stellar population reduction factor in $A_V$ of 0.44. We did not include an active galactic nuclei (AGN) model or the 2175 \AA\ bump in the fitting, and the nebular component is excluded (see Section~\ref{sec:phot}). The models also included a separation age between young and old stars of 10 Myr (default value for CIGALE). We defer discussions on our choices in SSP models and star formation histories to Section \ref{sec:model_reasons}.

\noindent \textbf{Physical Parameters.} The modelled parameters are listed in Table \ref{tab:CIGALE}.  Here we provide a brief discussion of the physical stellar population parameters and refer readers to \cite{CIGALE} for more information. The $t_{\rm age}$ parameter is the age of the oldest stars in the galaxy. The values for ${\rm \tau_{main}}$ and $t_{\rm age}$ were chosen to allow for a wide range (star-forming to passive) of possible star formation histories. The $A_V$ parameter measures the \textit{V}-band attenuation for the young stellar population used in the dust attenuation model. The values for $A_V$ were chosen to allow for a broad range of possible attenuation corrections, from zero reddening to three magnitudes.

\begin{figure}
  \includegraphics[width=\columnwidth,keepaspectratio]{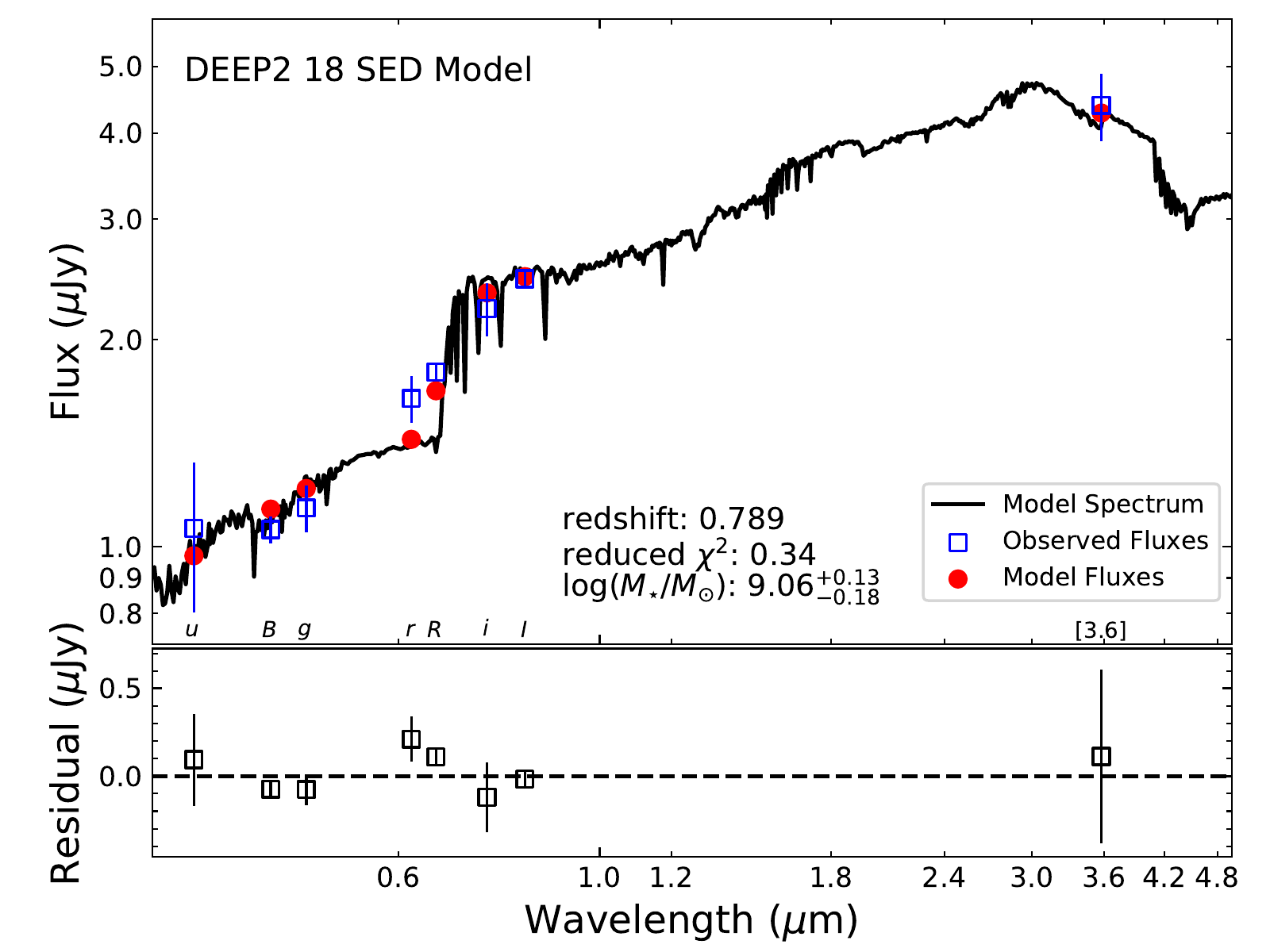}
  \vspace{-0.5cm}
  \caption{Top: The best-fit spectral synthesis model constructed by CIGALE for DEEP2 \#18. The rightmost blue square is the new \textit{Spitzer}/IRAC 3.6 $\micron$ photometry. Bottom: Difference between observed and model fluxes.}
  \label{fig:SED}
\end{figure}

Using these adopted parameters, CIGALE created thousands of models for each galaxy and determined the best fitting model using a Bayesian-like approach. A Bayesian methodology provides robust estimates on the physical properties, their associated uncertainties, and accounts for degeneracies between parameters by weighting each model by their $\chi^2$ \citep{Boquien19}. The results of the SED modelling are summarized in Table \ref{tab:CIGALE_results}. We illustrate in Figure \ref{fig:SED} the SED-fitting result from CIGALE for one of our galaxies, DEEP2 \#18.

We note that the SED results for DEEP2 \#04 and \#14 are not used. The SED models for these galaxies are not well constrained due to their limited photometric dataset. DEEP2 \#04 is a very bright galaxy that its optical measurements are not available in the DEEP2 catalogs. DEEP2 \#14 was not detected by \textit{Spitzer}/IRAC, thus limiting its photometric dataset to three bands (\textit{BRI}) and an upper limit at 3.6 $\mu$m. The upper limit is far below the \textit{BRI} photometry, causing a poor fit ($\chi^2_\nu = 5.7$).

\begin{figure}
	\includegraphics[width=\columnwidth,keepaspectratio]{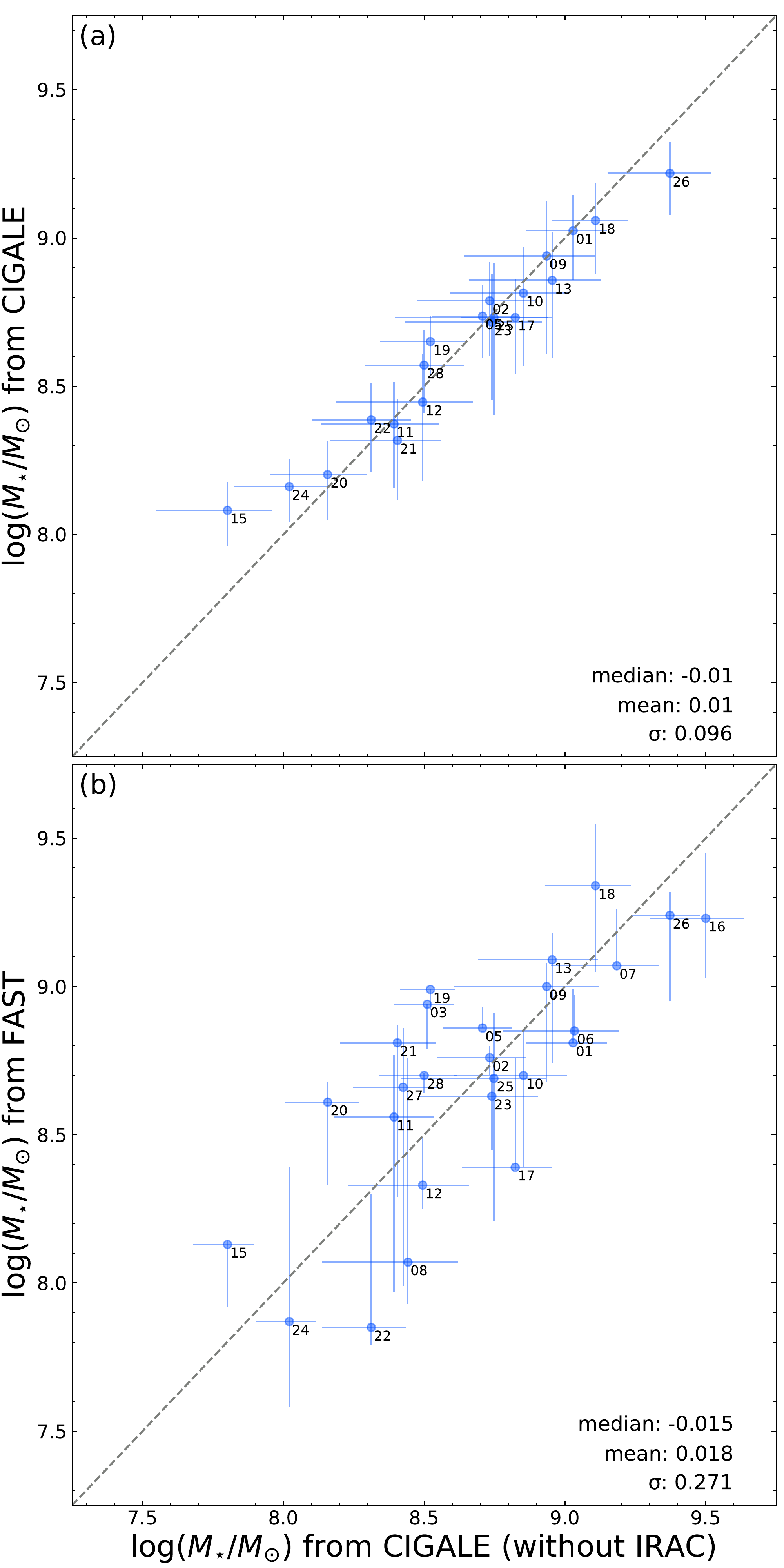}
    \vspace{-0.25cm}
    \caption{(a) Comparison between stellar masses derived from CIGALE with and without IRAC photometry. Blue circles are \z\ DEEP2 galaxies with IRAC detections (3.6 $\mu$m or 3.6 and 4.5 $\mu$m). Galaxies with confused measurements are excluded. Good agreement is seen between stellar masses derived with and without infrared photometry. (b) Comparison between stellar masses derived from CIGALE and FAST (see \citealt{Ly2015b}) without the use of \textit{Spitzer}/IRAC photometry. The grey dashed line marks one-to-one. The median, mean, and $\sigma$ values are calculated from the difference in mass between the two stellar mass estimates illustrated here. Generally, these mass estimates using different SED fitting code agree within measurement uncertainties.}
    \label{fig:mass_comparison}
\end{figure}

\subsubsection{Explanations for Assumptions in SED Fitting}
\label{sec:model_reasons}

There are two reasons why we chose the \citetalias{BC03} SSP model. First, it is more commonly used for galaxy evolution studies, which enable direct comparisons to earlier studies. Specifically, we will later compare our work against the local \MZ\ relation of \citetalias{AM13}, which used stellar masses determined with the \citetalias{BC03} SSP model \citep{Kauffmann03}.

Second, \citetalias{BC03} is the more conservative choice, as CIGALE only supports the \citetalias{BC03} and \citet[hereafter M05]{Maraston05} SSP models. \citetalias{Maraston05} was one of the first SSP models to include the thermally-pulsating asymptotic giant branch (TP-AGB) phase. TP-AGB stars heavily affect the SED of galaxies at stellar ages of about 1 to 3 Gyr. Its effects are still debated \citep[e.g.,][and references therein]{Kriek10,Capozzi16}; we refer readers to \cite{Conroy09}, which discussed differences in physical properties for these models.

We note that this age range is generally older than the ages of our galaxies, given our emission-line selection, thus TP-AGB stars are not expected to have a significant effect in our modelling. This is supported by the fact that stellar masses, which we derived with the \citetalias{Maraston05} SSP, are only systematically lower by a median (mean) value of 0.14 (0.13) dex compared to \citetalias{BC03} (assuming a \cite{Salpeter55} IMF for both models).

A delayed $\tau$ star formation history is well-suited for young galaxies with ages < $\tau$ and is commonly used for both star-forming and quiescent galaxies at high redshifts \citep[e.g.,][]{Papovich11,Pacifici13,Lee15,Lee18,Estrada19}. Furthermore, with our limited photometric dataset, other star formation histories, such as those with episodic bursts, have more degrees of freedom to constrain.

\subsubsection{Stellar Mass Comparisons}
\label{sec:mass_comp}
In panel (a) of Figure \ref{fig:mass_comparison}, we examine how the new \textit{Spitzer}/IRAC observations constrain the masses of the 28 galaxies. Here, we illustrate the masses derived from CIGALE including and excluding the new \textit{Spitzer}/IRAC observations. As illustrated, the stellar masses estimated with or without the new observations are consistent for most of the galaxies with a dispersion below 0.1 dex. This suggests that the other galaxies without \textit{Spitzer} measurements have robust stellar mass estimates from mostly optical data. In panel (b) of Figure \ref{fig:mass_comparison}, we compare stellar masses calculated from CIGALE and a different SED fitting code, FAST \citep{Kriek09}. These FAST estimates were derived by \cite{Ly2015b}. To compare directly, we do not include the new \textit{Spitzer}/IRAC 3.6 and 4.5 $\micron$ observations. While there are differences between the FAST and CIGALE estimates, there is generally good agreement with a dispersion of 0.27 dex. The good agreement between CIGALE and FAST indicates that we are able to conduct our \MZ\ analysis with the DEEP2 and \MACT\ samples.

In addition, we compare our stellar mass estimates against those obtained from modelling DEEP2 optical spectra \citep[hereafter C17]{Comparat17}. Here we briefly summarize the sample and modelling, and refer readers to \citetalias{Comparat17} for more details. From the fourth DEEP2 data release, $\approx$23,000 unique galaxy spectra were selected with a redshift between 0.7 and 1.2. SED models were created using the FIREFLY fitting code \citep{Wilkinson17}, which uses a \cite{Maraston11} SSP, three different stellar libraries, and three different IMFs.
In the below comparisons, we adopt a \cite{Salpeter55} IMF for all stellar mass estimates, and use the \citetalias{Comparat17} estimates that adopted the STELIB stellar library \citep{Stelib}.

Out of our 28 DEEP2 galaxies, 27 were modelled by \citetalias{Comparat17} (DEEP2 \#15 was not modelled).  For \citetalias{BC03}, we find that \citetalias{Comparat17} stellar masses were higher by a median (mean) value of 0.47 (0.44) dex. For \citetalias{Maraston05}, \citetalias{Comparat17} stellar masses were again higher by a median (mean) value of 0.62 (0.56) dex. The higher stellar masses measured by \citetalias{Comparat17} is possibly due to the inability to constrain the Balmer stellar absorption lines.  Specifically \citetalias{Comparat17} masked regions affected by nebular emission lines, which included the Balmer lines. The best fits revealed significant stellar absorption, which was not seen from double Gaussian fitting by \cite{Ly2015b}. We note that the average stellar mass uncertainty from \citetalias{Comparat17} is 0.62 dex. This uncertainty is larger than that obtained from modelling broadband optical and infrared data (0.17 dex). One of the factors contributing to the larger uncertainty is that obtaining significant detection of the continuum with spectroscopy is more difficult for these low-mass galaxies.

\subsection{$T_e$-based Oxygen Abundances}

Both \OIII 4363 and \OIII 5007 are collisionally excited emission lines where their flux ratio is dependent on the free electron kinetic temperature of the ionized gas, $T_e$. The cooling of the gas is controlled by collisional excitation of ionized atoms, such as single and doubly ionized oxygen, and radiation from these excited states. However, for metal-poor gas, cooling is difficult with the lack of ionized metals. Thus, the electron temperature of the gas is high and results in more collisions to the excited doubly-ionized oxygen state that produces the \OIII 4363 emission \citep{Aller84}.
The gas-phase metallicity for our galaxies was determined by \cite{Ly2015b} by following previous direct metallicity studies and using the empirical relations of \cite{Izotov06}. Here, we briefly summarize the approach, and we refer readers to \cite{Ly2014} for more details. The electron temperature is determined from the flux ratio of {\OIII$\lambda$4959,5007} against \OIII 4363. Using $T_e$ and two emission-line flux ratios, \OII$\lambda$3726,3729/H$\beta$ and \OIII$\lambda$4959,5007/H$\beta$, the oxygen-to-hydrogen ratio (O/H) can be determined.
A standard two-zone temperature model with $T_e$(O$^+$) = 0.7$T_e$(O$^{++}$) + 3000 K was adopted \citepalias{AM13}. The median, mean, and dispersion in metallicity for the DEEP2 galaxies are $12+\logOH = 8.17$, 8.1, and 0.26, respectively.

\begin{figure*}
  \includegraphics[width=7.0in,keepaspectratio]{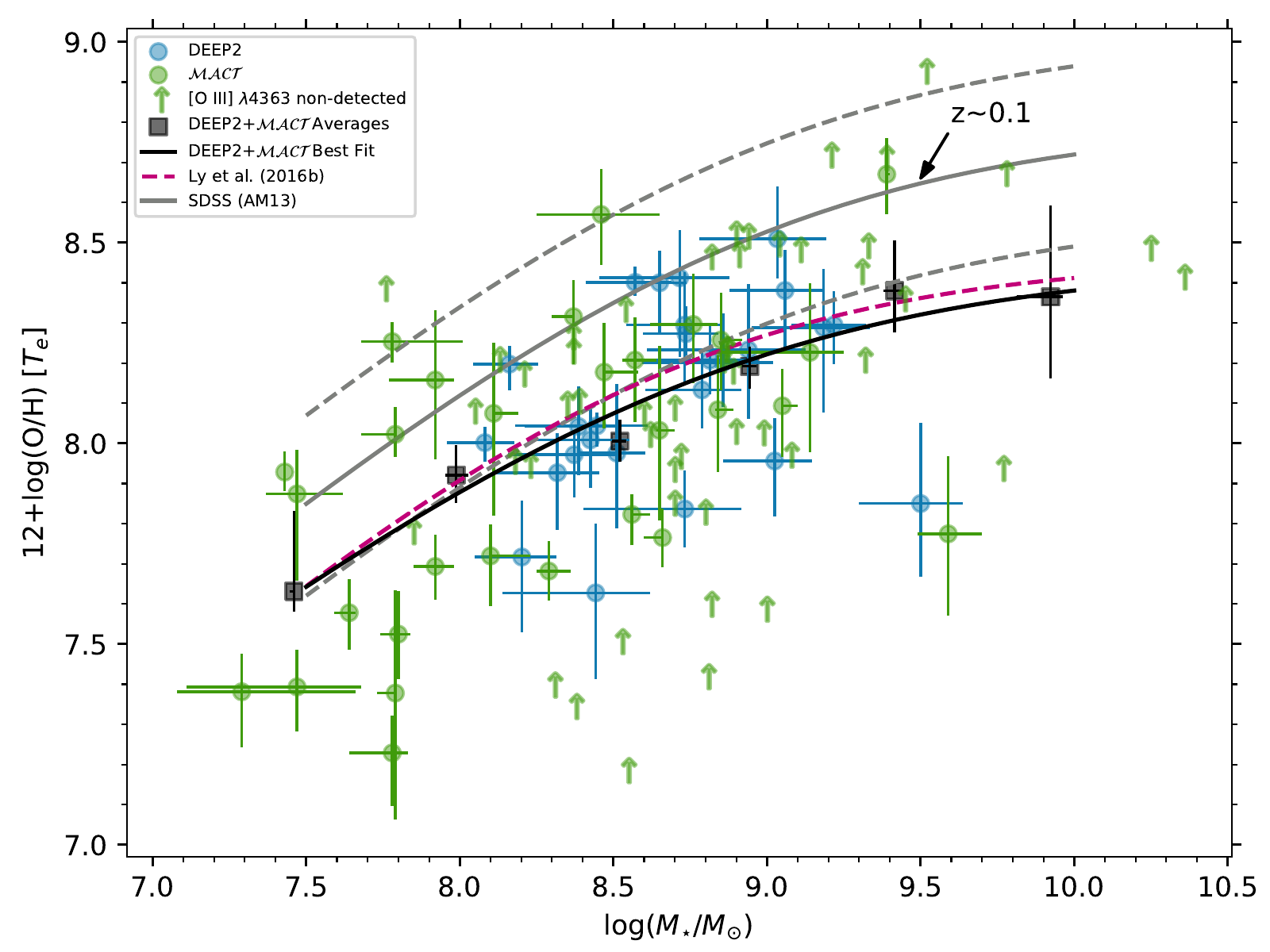}
  \vspace{-0.75cm}
  \caption{Oxygen abundance as a function of stellar mass. Blue circles are  \z\ DEEP2 galaxies from this study. Green circles and green arrows are the \MACT\ \OIII 4363-detected and non-detected samples, respectively. The DEEP2 and \MACT\ galaxies are binned in stellar mass (see Table \ref{tab:bin}), and the resulting average mass and metallicity of the bins are shown as black squares with their corresponding uncertainty. The binned points are fitted to Equation (\ref{equ_2}) (black line; see Figure \ref{contour} for fitting results). We compare our \MZ\ relation to SDSS galaxies from \protect\citetalias{AM13} (solid grey line, with grey dashed lines enclosing $\pm$1$\sigma$), and the DEEP2+\MACT\ relation from \protect\cite{Ly2016b} (dashed magenta line).}
	\label{fig:mass}
\end{figure*}

\section{Results}
\label{sec:4}

We describe below the stellar population of these low-metallicity galaxies (Section \ref{sec:stellar_pop}), measurements of the \MZ\ relation (Section \ref{sec:relation}), comparisons to previous studies (Section \ref{sec:Z13}), dependence on SFR for the \MZ\ relation (Section \ref{sec:SFR}), and the selection function of our study (Section \ref{sec:selection}).

\subsection{Stellar Population}
\label{sec:stellar_pop}

The 28 galaxies in our sample were originally classified as low-mass galaxies based on their optical photometry. However, infrared data is needed to observe the low-mass stars in these galaxies. Using the 3.6 $\micron$ (and 4.5 $\micron$ where available) photometry from \textit{Spitzer}/IRAC and existing optical data, our SED models have confirmed that the 28 galaxies have low masses, with stellar masses between \logM=8.08 and 9.50 with a median (mean) of 8.72 (8.69). The good agreement in stellar mass measurements with and without \textit{Spitzer} infrared data, as illustrated in Figure~\ref{fig:mass_comparison}(b), suggests that there is no evidence for an old ($\gtrsim$1 Gyr) population of low-mass stars in these galaxies. This result is driven primarily by the low luminosities observed at rest-frame 2 $\mu$m.

We note that we considered bottom-heavier IMFs by running CIGALE with a \cite{Salpeter55} IMF. As expected, the masses with the \cite{Salpeter55} IMF are systematically higher by 0.22 dex. While the stellar masses are higher, the results still support the lack of a significant old population of low-mass stars in these galaxies.

For our sample, we have spectroscopic measurements of H$\beta$ and higher-order Balmer lines (e.g., H$\gamma$, H$\delta$).  These measurements allow us to measure the SFRs using the H$\beta$ luminosity corrected for dust attenuation using the H$\gamma$/H$\beta$ Balmer decrement \citep{Ly2015b}. Combining the Balmer-derived SFRs with derived stellar masses, the specific SFR (sSFR $\equiv$ SFR/$M_{\star}$) for these galaxies range from $\log{\left({\rm sSFR}/{\rm yr}^{-1}\right)}$ = --8.79 to --6.62 with a median and mean of --8.1 and --8.0, respectively.

\subsection{\MZ\ Relation}
\label{sec:relation}

\begin{figure*}
  \centering
  \includegraphics[width=7.0in,keepaspectratio]{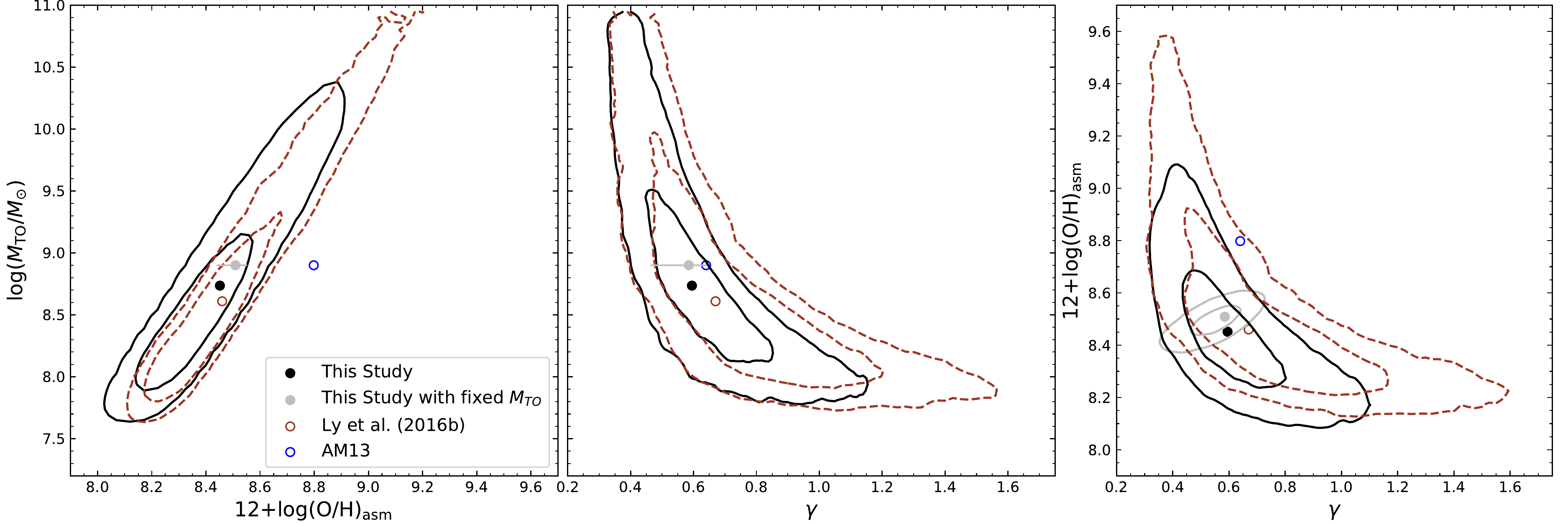}
  \vspace{-0.5cm}
  \caption{Confidence contours (68\% and 95\%) for the best fit parameters of the \MZ\ relation (see Table \ref{tab:best_fit}). Our fitting results for the \MZ\ relation are shown by solid black circles, while filled grey circles and grey contours are for $\log(\MTO/M_{\odot})$ fixed to 8.901 from \protect\citetalias{AM13}. Overlaid in brown dashed lines and unfilled brown circles are the fitting results from \protect\cite{Ly2016b}. Blue unfilled circles are the best fit parameters from \protect\citetalias{AM13}.}
    \label{contour}
\end{figure*}

In Figure \ref{fig:mass}, we examine the gas-phase metallicity dependence on stellar mass at \z\ for the 28 DEEP2 galaxies and the \MACT\ sample, and compare it against the $z\sim0.1$ \MZ\ relation from \protect\citetalias{AM13}. Here we use the following formalism from \cite{Moustakas11} to describe the shape of the \MZ\ relation:
\begin{equation}
  12 + \logOH = 12 + \logOH_{\rm asm} -\log\left[1+\left(\frac{\MTO}{M_{\star}}\right)^{\gamma}\right],
  \label{equ_2}
\end{equation}

\noindent where $\logOH_{\rm asm}$ is the asymptotic metallicity at the high-mass end, $\MTO$ is the turnover mass in the \MZ\ relation, and $\gamma$ is the power-law slope at the low-mass end. We fit this equation to the average mass and metallicity of the DEEP2 galaxies and the \MACT\ sample in each stellar mass bin. These averages are given in Table \ref{tab:bin} and are illustrated in Figure \ref{fig:mass} as black squares. In our fitting, we use a Monte Carlo approach where we repeat the fit  50\,000 times using uncertainties obtained from the bootstrap method. In our fitting of the \MZ\ relation, we allow $\MTO$ (1) to be a free parameter, and (2) fixed it to the local value from \citetalias{AM13}. The results are given in Table \ref{tab:best_fit}, and confidence contours are illustrated in Figure \ref{contour}.

Our best fit of the \MZ\ relation illustrates that there is a strong evolution between $z\sim0.1$ and \z. At a given stellar mass, the \z\ \MZ\ relation is lower than the relation from \citetalias{AM13} on average by 0.27 dex.
This evolution follows \zevo. The overall shape of the \MZ\ relation between \citetalias{AM13} and our results is similar, comprising of a steep rise in metallicity at low stellar masses then transitioning to a plateau at high stellar masses. Additionally, the overall shape of the \MZ\ relation is consistent with the relation found by \cite{Ly2016b}.

\subsection{Comparison with Previous \z\ \MZ\ Studies}
\label{sec:Z13}

Recent studies, such as \citet[hereafter Z13]{Zahid13} and \cite{Guo16}, have measured the evolution of the \MZ\ relation at intermediate redshifts. In these studies, strong-line metallicity diagnostics were used to determine chemical abundances. Specifically, \citetalias{Zahid13} used the \citet[hereafter KK04]{Kobulnicky04} calibration. Utilizing this calibration on our DEEP2 sample would yield metallicities that are 0.5--1 dex higher than those determined with the $T_e$ method. This suggests that the local \citetalias{Kobulnicky04} calibration is not valid for metal-poor galaxies at \z. The use of different metallicity diagnostics inhibits a direct comparison against our results.  However, relative comparisons against $z\approx0.1$ measurements using the same metallicity diagnostics can be examined.\footnote{We compare our results only to \citetalias{Zahid13} because \cite{Guo16} utilized a strong-line metallicity calibration that makes it difficult to conduct relative comparisons and because their results were more uncertain.}

In \citetalias{Zahid13}, a sample of 1254 galaxies at $0.75 < z < 0.82$ was selected from the third data release\footnote{\url{http://deep.ps.uci.edu/dr3/}} of the DEEP2 Galaxy Redshift Survey. Here, we briefly summarize the sample and refer readers to \citetalias{Zahid13} for more details. Among the $\sim$50,000 spectra, galaxies were selected with \OII3727 and H$\beta$ detections at S/N $>$ 3. In addition, 17 X-ray galaxies were removed to limit AGN contamination. Stellar masses were determined from \textit{BRI}-band photometry, and average Balmer stellar absorption was corrected for each of the selected galaxies.

\begin{figure}
  \includegraphics[width=\columnwidth,keepaspectratio]{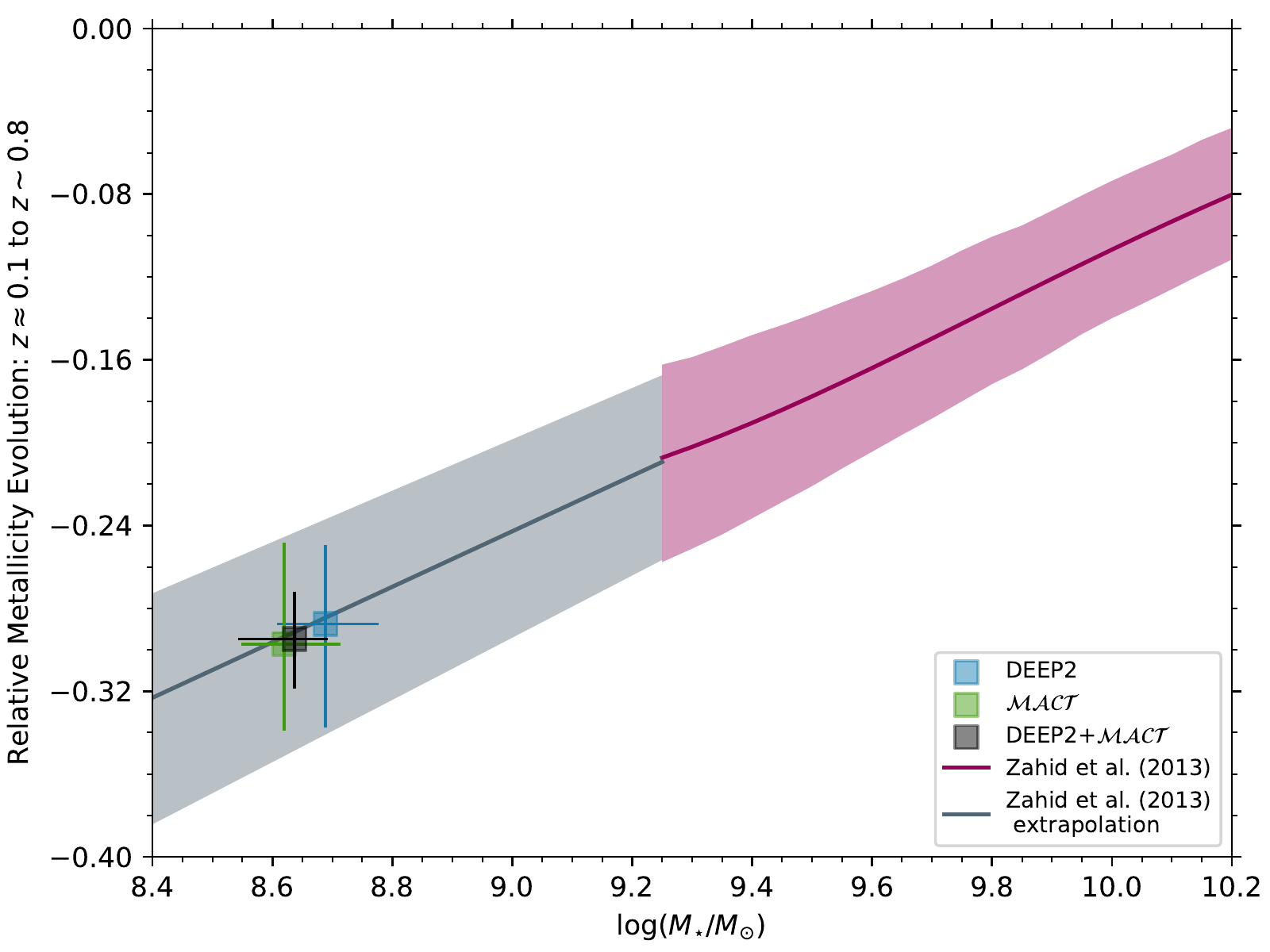}
  \vspace{-0.5cm}
  \caption{Relative metallicity evolution ($\Delta \log({\rm O/H})$) in the \MZ\ relation as a function of stellar mass. The blue, green, and black squares are the average $T_e$-based metallicity for DEEP2 galaxies from this study, \MACT\ galaxies, and DEEP2 and \MACT, respectively. We compare our results to \citetalias{Zahid13} (magenta line with shaded region enclosing $\pm$1$\sigma$), which used strong-line metallicity diagnostics. In both studies, relative evolution is derived from SDSS ($z\approx0.1$).  Additionally, we extrapolate \citetalias{Zahid13} results to lower stellar masses (grey line with shaded region enclosing $\pm$1$\sigma$). The evolution in our \MZ\ relation is in good agreement with the extrapolation of \citetalias{Zahid13} results toward lower stellar masses. This demonstrates that lower mass galaxies build up their interstellar metal content more recently.}
  \label{fig:MZ-evolution}
\end{figure}

In Figure \ref{fig:MZ-evolution}, we illustrate the evolution in metallicity as a function of stellar mass between $z\approx0.1$ and \z\ for \citetalias{Zahid13}. Additionally, we extrapolate \citetalias{Zahid13} results to lower stellar masses using a linear fit, $\Delta Z$ $\propto$ 0.13\,\logM, and compare it to the average evolution in metallicity of our samples (DEEP2, \MACT, and DEEP2+\MACT). Here we take the average metallicity of our samples due to a limited sample size and the effects of the selection function at high masses (see Section~\ref{sec:selection}). As shown in the figure, our evolution in metallicity is in good agreement with the extrapolated evolution of \citetalias{Zahid13} toward lower stellar masses. The difference in $\Delta Z$ between \citetalias{Zahid13} and our study is expected, because we are sampling two different stellar populations. For reference, \citetalias{Zahid13} galaxies are higher mass galaxies with stellar masses between \logM\ = 9.25 and 10.5, while our low-mass galaxies range from \logM\ = 8.08 to 9.50. As illustrated in Figure \ref{fig:MZ-evolution}, high-mass galaxies have accumulated most of their metallicity before \z. Between \z\ and $z\approx0.1$, the chemical evolution for \citetalias{Zahid13} is low, ranging from $\Delta Z$ = --0.21 to --0.08 dex. On the contrary, low-mass galaxies appear to undergo significant metal enrichment, with our samples (DEEP2, \MACT, and DEEP2+\MACT) at $\Delta Z$ = --0.29, --0.30, and --0.29 dex, respectively. Thus, a dwarf galaxy of $M_{\star} \approx 4\times10^8$ \Msun\ at \z\ undergoes 1.6 times more enrichment than a massive galaxy of $M_{\star} \approx 10^{10}$ \Msun\ at the same redshift.

As a further test, we can directly compare our evolution in the \MZ\ relation to \cite{Ly2016b}. We determined that the \MZ\ relation follows a \zevo\ evolution, such that metallicity is lower by $\approx$0.27 dex at \z\ compared to $z=0.1$. This evolution is in agreement, within measurement uncertainties, of \cite{Ly2016b}: $(1+z)^{-2.32^{+0.52}_{-0.26}}$. We note that the higher dependence measured by \cite{Ly2016b} is due to a higher (i.e., more enriched) \MZ\ relation at $z\approx0.07$ than \citetalias{AM13}.

\subsection{SFR Dependence}
\label{sec:SFR}

The idea that the \MZ\ relation has a secondary dependency on the SFR has been proposed by a number of observational and theoretical studies \citep[e.g.,][]{Ellison08, Lara10, Mannucci10, Dave11, Lilly13, Salim14}. 
Specifically, it has been suggested that higher SFRs can account for the lower metal abundances at higher redshifts, such that the \MZ\ relation is part of a larger non-evolving metallicity relation. To test this idea, we follow the approach of adopting a non-parametric method of projecting the \MZ--SFR relation onto two-dimensional spaces. This was first implemented by \cite{Salim14} and later applied to higher redshifts by \cite{Salim15} and \cite{Ly2016b}.

First, we consider how the \MZ\ relation depends on sSFR, which we illustrate in the left panel of Figure \ref{fig:bin_mass_SFR}. The \MZ\ relation is divided into two sSFR bins, at the median of our DEEP2+\MACT\ galaxies. Average measurements from composite spectra of \citetalias{AM13} are overlaid as grey squares. For reference, the local \MZ\ relation of \citetalias{AM13} is shown by the grey line. At a given stellar mass, the left panel of Figure \ref{fig:bin_mass_SFR} suggests that the metallicity for \z\ low-mass galaxies does not decrease as the (s)SFR increases.

Next, we consider how the metallicity and sSFR depend on mass. This is illustrated in the right panel of Figure \ref{fig:bin_mass_SFR}. The galaxies are divided into two stellar mass bins, at the median of our samples.
Galaxies from \citetalias{AM13} are overlaid as grey squares. In each stellar mass bin, the galaxies from \citetalias{AM13} demonstrate that there is a moderate negative correlation between metallicity and sSFR, as expected. However, the metallicities of the DEEP2+\MACT\ galaxies do not decrease as the sSFRs increase. In the lower mass bin, there are a number of low-sSFR and low-metallicity\footnote{\OIII 4363 is detected.} galaxies ($12+\logOH \sim 7.5$), which go against what is predicted by \cite{Mannucci10}.

Finally, we consider how the $M_{\star}$--SFR relation depends on metallicity. This is illustrated in the left panel of Figure \ref{fig:bin_met}. The galaxies are divided into four $12+\logOH$ bins ranging from 7.19 to 9.09.
Galaxies from \citetalias{AM13} are overlaid as grey squares.
The right panel of Figure \ref{fig:bin_met} illustrates the sSFR dependence on metallicity. The average values of sSFR, illustrated in the left panel, are overlaid as black squares in the right panel. We fit our DEEP2+\MACT\ \OIII 43636-detected galaxies to a linear relation using \texttt{scipy.optimize.curve\_fit}. This was repeated 40\,000 times, perturbing the galaxies in metallicity and sSFR to account for measurement uncertainties.  We find that the best fit is
\begin{equation}
\log({\rm sSFR}) = -6.69^{+0.68}_{-0.76} - (0.17^{+0.09}_{-0.08}) \cdot [12+\logOH]. 
\end{equation}
For comparison, we run the same fitting analysis on the stacked results from \citetalias{AM13}, and find a best fit of 
\begin{equation}
\log({\rm sSFR}) = 10.24^{+0.17}_{-0.19} - (2.27\pm0.02) \cdot [12+\logOH].
\end{equation}
The shallower slope that we measure is due to (1) the selection function missing metal-rich low-sSFR galaxies (see Section~\ref{sec:selection}), and (2) a population of extremely low metallicity galaxies that have relatively low sSFR. We note that the latter is not due to a selection bias since \OIII 4363 is more easily detected in low-metallicity, high-sSFR galaxies.

Our best fit yields $t=-2.04$ and $p=0.046$ for a linear regression t-test. Thus, it suggests that we cannot rule out the null hypothesis (i.e., no correlation) at $>$95\% confidence. We note that a t-test of \citetalias{AM13} sSFR--$Z$ relation with our DEEP2+\MACT\ data yields $t=25.5$ with $p < 10^{-5}$ ($>$4$\sigma$).  This suggests that we are able to rule out a strong sSFR--$Z$ dependence.

In summary, we conclude that the \MZ\ relation at \z\ does not have a significant secondary dependence on SFR such that galaxies with higher sSFR have reduced metallicity. As illustrated in Figures \ref{fig:bin_mass_SFR} and \ref{fig:bin_met}, in each of the three projections, our galaxies display either a mild dependence or no dependence. Our result is in agreement with previous studies that have found a moderate or no SFR dependence of the \MZ\ relation \citep[e.g.,][]{Perez13, Reyes15, Salim15, Guo16, Sanchez17}. However, several studies have found a strong SFR dependence for galaxies in the local Universe \citep{Lara10, Lilly13, Salim14}, and for high redshift galaxies, studies have found a similar trend but offset from the local relation \citep[i.e. there is redshift dependence;][]{Zahid14, Salim15}.

\subsection{\OIII 4363 Selection Function}
\label{sec:selection}
The requirement to detect \OIII 4363 in our spectroscopic study leads to a concern of selection bias for our samples. Here, we briefly outline the effects on our results. We refer readers to \cite{Ly2016b} for an in-depth analysis.

The detection of the \OIII 4363 line is dependent on a combination of (1) the electron temperature/metallicity of the gas \citep{Nicholls14}, and (2) the dust-corrected SFRs which determine the absolute strengths of the emission-line fluxes.
At high SFRs, \OIII 4363 is detectable for a wide range of metallicities. As SFR decreases, only metal-poor galaxies can be detected in an emission-line flux-limited survey. The effects of the \OIII 4363 selection function is illustrated in the right panel of Figure \ref{fig:bin_met} where sSFR decreases as metallicity increases, but at a lower rate than \citetalias{AM13}. The difference in the slopes is the result of two factors at high and low metallicities. First, there is a selection toward metal-rich galaxies with high SFRs because those with low SFRs will not have \OIII 4363 detections and will fall below the flux limit cuts adopted by \cite{Ly2015b}. Second, the DEEP2+\MACT\ sample is lacking low-metallicity, high-sSFR galaxies. Unlike their metal-rich counterparts, these galaxies have detectable \OIII 4363 measurements.

\section{Conclusions}
\label{sec:5}
We present new \textit{Spitzer}/IRAC 3.6 $\micron$ (4.5 $\micron$ where available) photometry for 28 metal-poor strongly star-forming galaxies at \z\ from the DEEP2 Survey. These new measurements were necessary to better constrain the stellar masses of these galaxies by observing the light from low-mass stars. Our SED modelling of 0.35--3.6(4.5) $\mu$m data confirmed that these galaxies have low stellar masses, $M_{\star}\approx10^{8.1}$--10$^{9.5}$ \Msun.

Combining these stellar masses with metal abundances derived from the electron temperature method, and measurements from the \MACT\ Survey, we determined that the \MZ\ relation evolves toward lower metallicity by $\approx$0.27 dex between \z\ and $z=0.1$: $\propto$\zevo. Our evolution is in agreement within measurement uncertainties of \cite{Ly2016b}: $\propto(1+z)^{-2.32^{+0.52}_{-0.26}}$. We compared the evolution in metallicity from \z\ to $z\approx0.1$ against recent studies at intermediate redshifts. Specifically, we find good agreement with those reported by \citetalias{Zahid13} when extrapolated toward lower stellar masses. This suggests that lower mass galaxies ($4\times10^8$ \Msun) chemically enriched their interstellar contents 1.6 times more rapidly than high mass galaxies ($10^{10}$ \Msun). Additionally, we measured the shape of the \MZ\ relation at \z, which is consistent with the shape of local relation \citepalias{AM13}. Finally, we considered whether the \MZ\ relation has a secondary dependence on specific SFR. We determined an sSFR--$Z$ relation of: $\log({\rm sSFR}) = -6.69^{+0.68}_{-0.76} - (0.17^{+0.09}_{-0.08}) \cdot [12+\logOH]$. A linear t-test determined that we cannot rule out no dependence at $>$95\% confidence. However, a strong dependence, as determined for local galaxies \citepalias{AM13}, can be ruled out at $>$4$\sigma$ confidence.
 
\section*{Acknowledgements}
We thank the anonymous referee for providing constructive feedback that improved the paper.
This work is based on observations made with the {\it Spitzer Space Telescope}, which is operated by the Jet Propulsion Laboratory, California Institute of Technology under a contract with NASA. Support for this work was provided by NASA through an award issued by JPL/Caltech under grant NNN12AA01C, and through a NASA traineeship grant awarded to the Arizona/NASA Space Grant Consortium.
This research made use of Astropy,\footnote{\url{http://www.astropy.org}.} a community-developed core Python package for Astronomy \citep{Astropy}.
This study makes use of data from the NEWFIRM Medium-Band Survey, a multi-wavelength survey conducted with the NEWFIRM instrument at the KPNO, supported in part by the NSF and NASA.






\renewcommand*{\thefootnote}{\alph{footnote}}

\newcommand{\FBe}{\tnotex{tn:3a}}
\begin{table*}
  \centering
  \caption{Optical and Near-Infrared Photometry for \OIII 4363-detected DEEP2 Galaxies}
  \label{tab:phot}
  \begin{threeparttable}
    \begin{tabular}{ccccccccccccccccccccccccc}
      \hline\hline
      ID & $u$ & $B$ & $g$ & $r$ & $R$ & $i$ & $I$ & $z$ & $J$ & $H$ & $K_s$\\
      (1)& (2) & (3) & (4) & (5) & (6) & (7) & (8) & (9) & (10)& (11)& (12)\\\hline
01 & 1.58 & 1.88 & 1.94 & 2.04 & 2.16 & 3.02 & 2.96 & 3.27 & 4.26 & 3.37 & 3.21 \\ [-1pt]
   & $\pm$0.02 & $\pm$0.05 & $\pm$0.10 & $\pm$0.01 & $\pm$0.04 & $\pm$0.02 & $\pm$0.07 & $\pm$0.05 & $\pm$0.20 & $\pm$0.18 & $\pm$0.18 \\ [-1pt]
02 & 0.83 & 0.93 & 0.94 & 1.10 & 1.01 & 1.79 & 1.62 & 1.48 & \ldots & \ldots & \ldots \\ [-1pt]
   & $\pm$0.02 & $\pm$0.08 & $\pm$0.01 & $\pm$0.02 & $\pm$0.05 & $\pm$0.02 & $\pm$0.12 & $\pm$0.06 & \ldots & \ldots & \ldots \\ [-1pt]
03 & 0.44 & 0.76 & 0.67 & 0.79 & 0.75 & 1.04 & 1.06 & 1.29 & \ldots & \ldots & \ldots \\ [-1pt]
   & $\pm$0.01 & $\pm$0.04 & $\pm$0.01 & $\pm$0.01 & $\pm$0.03 & $\pm$0.02 & $\pm$0.05 & $\pm$0.05 & \ldots & \ldots & \ldots \\ [-1pt]
04 & \ldots & \ldots & \ldots & \ldots & \ldots & \ldots & \ldots & \ldots & \ldots & \ldots & 23.11 \\ [-1pt]
   & \ldots & \ldots & \ldots & \ldots & \ldots & \ldots & \ldots & \ldots & \ldots & \ldots & $\pm$0.15 \\ [-1pt]
05 & 1.47 & 1.55 & 1.73 & 1.87 & 1.80 & 2.23 & 1.27\FBe & 0.35\FBe & \ldots & \ldots & \ldots \\ [-1pt]
   & $\pm$0.06 & $\pm$0.04 & $\pm$0.04 & $\pm$0.07 & $\pm$0.08 & $\pm$0.11 & $\pm$0.18 & $\pm$0.17 & \ldots & \ldots & \ldots \\ [-1pt]
06 & \ldots & 1.58 & \ldots & \ldots & 2.23 & \ldots & 3.13 & \ldots & \ldots & \ldots & \ldots \\ [-1pt]
   & \ldots & $\pm$0.03 & \ldots & \ldots & $\pm$0.06 & \ldots & $\pm$0.15 & \ldots & \ldots & \ldots & \ldots \\ [-1pt]
07 & 2.05 & 2.74 & 3.03 & 2.90 & 3.50 & 4.07 & 5.02 & 1.20\FBe & \ldots & \ldots & \ldots \\ [-1pt]
   & $\pm$0.81 & $\pm$0.03 & $\pm$0.33 & $\pm$0.45 & $\pm$0.05 & $\pm$0.70 & $\pm$0.12 & $\pm$1.87 & \ldots & \ldots & \ldots \\ [-1pt]
08 & \ldots & 0.77 & \ldots & \ldots & 0.91 & \ldots & 1.10 & \ldots & \ldots & \ldots & \ldots \\ [-1pt]
   & \ldots & $\pm$0.03 & \ldots & \ldots & $\pm$0.05 & \ldots & $\pm$0.13 & \ldots & \ldots & \ldots & \ldots \\ [-1pt]
09 & \ldots & 0.93 & \ldots & \ldots & 1.21 & \ldots & 1.90 & \ldots & \ldots & \ldots & \ldots \\ [-1pt]
   & \ldots & $\pm$0.03 & \ldots & \ldots & $\pm$0.04 & \ldots & $\pm$0.10 & \ldots & \ldots & \ldots & \ldots \\ [-1pt]
10 & \ldots & 1.44 & \ldots & \ldots & 1.73 & \ldots & 2.30 & \ldots & \ldots & \ldots & \ldots \\ [-1pt]
   & \ldots & $\pm$0.03 & \ldots & \ldots & $\pm$0.04 & \ldots & $\pm$0.09 & \ldots & \ldots & \ldots & \ldots \\ [-1pt]
11 & \ldots & 0.87 & \ldots & \ldots & 0.96 & \ldots & 1.09 & \ldots & \ldots & \ldots & \ldots \\ [-1pt]
   & \ldots & $\pm$0.03 & \ldots & \ldots & $\pm$0.04 & \ldots & $\pm$0.11 & \ldots & \ldots & \ldots & \ldots \\ [-1pt]
12 & \ldots & 0.76 & \ldots & \ldots & 0.86 & \ldots & 1.20 & \ldots & \ldots & \ldots & \ldots \\ [-1pt]
   & \ldots & $\pm$0.03 & \ldots & \ldots & $\pm$0.03 & \ldots & $\pm$0.08 & \ldots & \ldots & \ldots & \ldots \\ [-1pt]
13 & \ldots & 0.75 & \ldots & \ldots & 1.00 & \ldots & 1.69 & \ldots & \ldots & \ldots & \ldots \\ [-1pt]
   & \ldots & $\pm$0.04 & \ldots & \ldots & $\pm$0.04 & \ldots & $\pm$0.10 & \ldots & \ldots & \ldots & \ldots \\ [-1pt]
14 & \ldots & 1.33 & \ldots & \ldots & 1.66 & \ldots & 2.05 & \ldots & \ldots & \ldots & \ldots \\ [-1pt]
   & \ldots & $\pm$0.04 & \ldots & \ldots & $\pm$0.05 & \ldots & $\pm$0.10 & \ldots & \ldots & \ldots & \ldots \\ [-1pt]
15 & 1.02 & 1.38 & 1.16 & 1.19 & 0.91 & 0.55\FBe & 0.86 & \ldots & \ldots & \ldots & \ldots \\ [-1pt]
   & $\pm$0.29 & $\pm$0.05 & $\pm$0.1 & $\pm$0.14 & $\pm$0.07 & $\pm$0.22 & $\pm$0.07 & \ldots & \ldots & \ldots & \ldots \\
16 & 0.67 & 1.08 & 1.07 & 1.63 & 2.08 & 3.17 & 3.38 & 1.94\FBe & \ldots & \ldots & \ldots \\ [-1pt]
   & $\pm$0.29 & $\pm$0.05 & $\pm$0.09 & $\pm$0.14 & $\pm$0.05 & $\pm$0.22 & $\pm$0.06 & $\pm$1.06 & \ldots & \ldots & \ldots \\ [-1pt]
17 & 0.8 & 1.6 & 1.42 & 1.19 & 2.01 & 1.98 & 2.40 & 2.36 & \ldots & \ldots & \ldots \\ [-1pt]
   & $\pm$0.27 & $\pm$0.06 & $\pm$0.09 & $\pm$0.13 & $\pm$0.07 & $\pm$0.20 & $\pm$0.06 & $\pm$0.98 & \ldots & \ldots & \ldots \\
18 & 1.06 & 1.06 & 1.14 & 1.64 & 1.80 & 2.22 & 2.45 & 1.66\FBe & \ldots & \ldots & \ldots \\ [-1pt]
   & $\pm$0.26 & $\pm$0.05 & $\pm$0.09 & $\pm$0.13 & $\pm$0.05 & $\pm$0.2 & $\pm$0.06 & $\pm$0.95 & \ldots & \ldots & \ldots \\ [-1pt]
19 & 1.6 & 1.69 & 2.13 & 2.10 & 1.85 & 2.23 & 1.87 & 2.04 & \ldots & \ldots & \ldots \\ [-1pt]
   & $\pm$0.32 & $\pm$0.04 & $\pm$0.11 & $\pm$0.16 & $\pm$0.08 & $\pm$0.25 & $\pm$0.07 & $\pm$1.17 & \ldots & \ldots & \ldots \\ [-1pt]
20 & 0.91 & 0.94 & 0.73 & 0.92 & 0.98 & 0.63 & 0.96 & 1.75\FBe & \ldots & \ldots & \ldots \\ [-1pt]
   & $\pm$0.28 & $\pm$0.05 & $\pm$0.09 & $\pm$0.13 & $\pm$0.07 & $\pm$0.21 & $\pm$0.07 & $\pm$1.0 & \ldots & \ldots & \ldots \\ [-1pt]
21 & 1.08 & 0.7 & 0.74 & 0.69 & 0.90 & 1.13 & 0.94 & 0.22\FBe & \ldots & \ldots & \ldots \\ [-1pt]
   & $\pm$0.29 & $\pm$0.05 & $\pm$0.10 & $\pm$0.14 & $\pm$0.05 & $\pm$0.23 & $\pm$0.09 & $\pm$1.22 & \ldots & \ldots & \ldots \\ [-1pt]
22 & 1.06 & 0.94 & 0.79 & 0.80 & 0.89 & 1.37 & 1.00 & 1.87 & \ldots & \ldots & \ldots \\ [-1pt]
   & $\pm$0.25 & $\pm$0.07 & $\pm$0.08 & $\pm$0.11 & $\pm$0.04 & $\pm$0.18 & $\pm$0.06 & $\pm$0.84 & \ldots & \ldots & \ldots \\ [-1pt]
23 & 0.93 & 0.84 & 0.69 & 0.92 & 0.94 & 1.73 & 1.34 & 1.97 & \ldots & \ldots & \ldots \\ [-1pt]
   & $\pm$0.22 & $\pm$0.07 & $\pm$0.09 & $\pm$0.12 & $\pm$0.04 & $\pm$0.21 & $\pm$0.07 & $\pm$0.69 & \ldots & \ldots & \ldots \\ [-1pt]
24 & 1.41 & 1.55 & 1.52 & 1.01 & 1.29 & 1.39 & 1.18 & 0.60\FBe & \ldots & \ldots & \ldots \\ [-1pt]
   & $\pm$0.24 & $\pm$0.07 & $\pm$0.10 & $\pm$0.13 & $\pm$0.04 & $\pm$0.23 & $\pm$0.06 & $\pm$0.76 & \ldots & \ldots & \ldots \\ [-1pt]
25 & 0.81 & 0.8 & 0.59 & 0.57 & 0.90 & 0.99 & 1.28 & 0.56\FBe & \ldots & \ldots & \ldots \\ [-1pt]
   & $\pm$0.21 & $\pm$0.14 & $\pm$0.09 & $\pm$0.12 & $\pm$0.05 & $\pm$0.21 & $\pm$0.08 & $\pm$0.68 & \ldots & \ldots & \ldots \\ [-1pt]
26 & \ldots & 3.08 & \ldots & \ldots & 4.44 & \ldots & 5.90 & \ldots & \ldots & \ldots & \ldots \\ [-1pt]
   & \ldots & $\pm$0.1 & \ldots & \ldots & $\pm$0.06 & \ldots & $\pm$0.08 & \ldots & \ldots & \ldots & \ldots \\ [-1pt]
27 & 0.77 & 0.94 & 0.90 & 1.06 & 1.14 & 1.42 & 1.24 & 1.62\FBe & \ldots & \ldots & \ldots \\ [-1pt]
   & $\pm$0.23 & $\pm$0.03 & $\pm$0.09 & $\pm$0.13 & $\pm$0.04 & $\pm$0.23 & $\pm$0.06 & $\pm$0.73 & \ldots & \ldots & \ldots \\ [-1pt]
28 & 2.63 & 2.58 & 2.48 & 2.56 & 2.26 & 2.75 & 1.98\FBe & 1.75\FBe & \ldots & \ldots & \ldots \\ [-1pt]
   & $\pm$0.24 & $\pm$0.03 & $\pm$0.09 & $\pm$0.13 & $\pm$0.04 & $\pm$0.22 & $\pm$0.07 & $\pm$0.73 & \ldots & \ldots & \ldots \\\hline
   \end{tabular}
    \begin{tablenotes}
    \item \leavevmode\kern-\scriptspace\kern-\labelsep (1): DEEP2 \OIII 4363 galaxy ID. (2)--(12): Photometric fluxes from shorter to longer wavelengths. These photometric data are obtained from several studies: $BRI$ \citep{Coil04}, $ugriz$ \citep[SDSS or CFHT;][]{Matthews13}, $JHK_s$ \citep[for \#01;][]{Whitaker11}, and $K_s$ \citep[for \#04;][]{Bundy06}. All fluxes are given in $\mu$Jy and have been corrected for nebular emission lines from optical spectroscopy using the technique described in \cite{Ly2014,Ly2016a}.
    \end{tablenotes}
    \begin{tablenotes}
    \item[a] \label{tn:3a} Fluxes are heavily affected by nebular emission line correction and not used in SED modelling.
    \end{tablenotes}
  \end{threeparttable}
\end{table*}

\begin{table*}
  \centering
  \caption{Adopted Parameters for SED Fitting with CIGALE}
  \label{tab:CIGALE}
  \begin{threeparttable}
    \begin{tabular}{cc}
      \hline\hline
      Parameter & Values \\
      (1) & (2)\\
      \hline
      $\rm{\tau_{main}}$ (Myr) & 250,500,1000,2000,4000,6000,8000\\
      $t_{\rm age}$ (Myr) & 10,25,50,75,100,125,150,175,200,225,250,500,1000,2000,4000,8000\\
      $A_V$(young) & 0.0,0.5,1.0,1.5,2.0,2.5,3.0\\
      \hline
    \end{tabular}
    \begin{tablenotes}
    \item \leavevmode\kern-\scriptspace\kern-\labelsep (1): $\tau_{\rm main}$ is the e-folding time of the main stellar population model, $t_{\rm age}$ refers to the age of the oldest stars in the galaxy, and $A_V$(young) is the $V$-band attenuation of the young population. All other parameters  were the default values set by CIGALE. (2): Allowed values for the SED modeling grid.
    \end{tablenotes}
  \end{threeparttable}
\end{table*}

\newcommand{\FBb}{\tnotex{tn:2a}}
\newcommand{\FBd}{\tnotex{tn:2b}}
\begin{table*}
  \centering
  \caption{Summary of SED-derived Stellar Properties from CIGALE}
  \label{tab:CIGALE_results}
  \begin{threeparttable}
    \begin{tabular}{lccccccccc}
      \hline\hline
    ID  & $\log(M_{\rm \star}/M_{\rm \odot})$ & SFR & SFR(10Myr) & SFR(100Myr) & $A_{\rm FUV}$ & $A_{\rm{V}}$ & $\tau_{\rm main}$ & Age & $\chi^{2}_{\nu}$\\
        & (dex) & ($M_{\rm \odot}$/yr) & ($M_{\rm \odot}$/yr) & ($M_{\rm \odot}$/yr) & (mag) & (mag) & (Gyr) & (Myr) & \\
    (1) & (2) & (3) & (4) & (5) & (6) & (7) & (8) & (9) & (10)\\\hline
    01 & $9.02^{+0.12}_{-0.17}$ & 9.66$\pm$8.31 & 9.27$\pm$5.66 & 7.52$\pm$4.05 & 1.52$\pm$0.54 & 0.53$\pm$0.21 & 3.14$\pm$2.81 & 390$\pm$388 & 1.06\\ [3pt]
    02 & $8.79^{+0.13}_{-0.19}$ & 2.45$\pm$1.64 & 2.41$\pm$1.48 & 2.13$\pm$1.04 & 0.91$\pm$0.72 & 0.31$\pm$0.25 & 3.38$\pm$2.79 & 915$\pm$736 & 1.71\\ [3pt]
    03 & $8.51^{+0.09}_{-0.12}$ & 12.17$\pm$11.2 & 9.90$\pm$6.49 & 6.60$\pm$5.35 & 3.39$\pm$0.80 & 1.35$\pm$0.38 & 3.18$\pm$2.77 & \ldots\FBb & 1.75\\ [3pt]
    04 & \ldots & \ldots & \ldots & \ldots & \ldots & \ldots & \ldots & \ldots & \ldots\\ [3pt]
    05 & $8.74^{+0.11}_{-0.14}$ & 11.24$\pm$7.49 & 10.43$\pm$5.78 & 6.97$\pm$3.31 & 1.94$\pm$0.44 & 0.72$\pm$0.20 & 3.20$\pm$2.77 & 156$\pm$155 & 0.1 \\ [3pt]
    06 & $9.03^{+0.16}_{-0.25}$ & 25.19$\pm$32.21 & 21.09$\pm$20.43 & 14.54$\pm$15.60 & 2.78$\pm$1.24 & 1.08$\pm$0.54 & 3.25$\pm$2.78 & \ldots\FBb & 0.03 \\ [3pt]
    07 & $9.18^{+0.15}_{-0.23}$ & 34.72$\pm$39.05 & 29.90$\pm$25.25 & 20.33$\pm$18.67 & 2.44$\pm$1.01 & 0.93$\pm$0.44 & 3.22$\pm$2.78 & \ldots\FBb & 0.25\\ [3pt]
    08 & $8.44^{+0.18}_{-0.30}$ & 7.53$\pm$9.24 & 6.06$\pm$5.38 & 4.31$\pm$4.46 & 1.90$\pm$1.02 & 0.72$\pm$0.43 & 3.20$\pm$2.78 & \ldots\FBb &0.01\\ [3pt]
    09 & $8.94^{+0.19}_{-0.33}$ & 12.91$\pm$13.26 & 11.83$\pm$8.83 & 8.17$\pm$6.31 & 2.68$\pm$0.91 & 1.02$\pm$0.39 & 3.28$\pm$2.77 & \ldots\FBb & 0.28\\ [3pt]
    10 & $8.81^{+0.16}_{-0.25}$ & 7.44$\pm$10.12 & 6.82$\pm$5.81 & 5.11$\pm$4.90 & 1.29$\pm$0.73 & 0.46$\pm$0.29 & 3.23$\pm$2.79 & \ldots\FBb & 0.23\\ [3pt]
    11 & $8.37^{+0.14}_{-0.21}$ & 4.78$\pm$5.52 & 4.25$\pm$3.21 & 2.93$\pm$2.62 & 1.27$\pm$0.66 & 0.46$\pm$0.27 & 3.16$\pm$2.78 & \ldots\FBb & 0.13\\ [3pt]
    12 & $8.45^{+0.16}_{-0.27}$ & 4.00$\pm$7.50 & 3.36$\pm$3.87 & 2.57$\pm$3.66 & 1.29$\pm$0.84 & 0.47$\pm$0.34 & 3.24$\pm$2.79 & \ldots\FBb & 0.05\\ [3pt]
    13 & $8.86^{+0.16}_{-0.26}$ & 7.87$\pm$21.38 & 5.85$\pm$10.66 & 5.08$\pm$10.53 & 1.68$\pm$1.07 & 0.60$\pm$0.44 & 3.24$\pm$2.79 & \ldots\FBb &0.15\\ [3pt]
    14\FBd & \ldots & \ldots & \ldots & \ldots & \ldots & \ldots & \ldots & \ldots & \ldots\\ [3pt]
    15 & $8.08^{+0.10}_{-0.12}$ & 3.81$\pm$1.69 & 3.51$\pm$1.37 & 2.05$\pm$0.77 & 0.79$\pm$0.54 & 0.29$\pm$0.21 & 3.16$\pm$2.77 & 90$\pm$52 & 2.65\\ [3pt]
    16 & $9.50^{+0.14}_{-0.20}$ & 33.44$\pm$70.79 & 27.53$\pm$36.26 & 23.22$\pm$34.89 & 3.66$\pm$1.09 & 1.39$\pm$0.50 & 3.27$\pm$2.79 & \ldots\FBb & 0.13\\ [3pt]
    17 & $8.73^{+0.13}_{-0.19}$ & 19.18$\pm$35.97 & 11.28$\pm$17.46 & 10.63$\pm$17.61 & 1.41$\pm$1.39 & 0.53$\pm$0.59 & 3.05$\pm$2.76 & 369$\pm$348 & 1.69\\ [3pt]
    18 & $9.06^{+0.13}_{-0.18}$ & 26.50$\pm$52.20 & 19.62$\pm$25.67 & 15.99$\pm$25.48 & 2.94$\pm$0.96 & 1.11$\pm$0.43 & 3.17$\pm$2.78 & \ldots\FBb & 0.34\\ [3pt]
    19 & $8.65^{+0.09}_{-0.11}$ & 15.70$\pm$6.49 & 14.27$\pm$5.08 & 8.35$\pm$2.89 & 1.79$\pm$0.46 & 0.68$\pm$0.19 & 3.17$\pm$2.77 &84$\pm$57 & 0.93\\ [3pt]
    20 & $8.20^{+0.11}_{-0.15}$ & 4.12$\pm$3.17 & 3.70$\pm$2.19 & 2.40$\pm$1.46 & 1.13$\pm$0.56 & 0.42$\pm$0.23 & 3.20$\pm$2.78 & 135$\pm$109 & 0.84\\ [3pt]
    21 & $8.32^{+0.14}_{-0.20}$ & 4.46$\pm$7.12 & 3.64$\pm$3.70 & 2.74$\pm$3.44 & 1.06$\pm$0.72 & 0.39$\pm$0.30 & 3.16$\pm$2.78 & 225$\pm$218 & 0.63\\ [3pt]
    22 & $8.39^{+0.12}_{-0.17}$ & 5.33$\pm$3.09 & 4.97$\pm$2.47 & 3.15$\pm$1.37 & 1.69$\pm$0.58 & 0.63$\pm$0.24 & 3.20$\pm$2.78 & \ldots\FBb & 0.75\\ [3pt]
    23 & $8.72^{+0.16}_{-0.26}$ & 3.83$\pm$6.02 & 3.52$\pm$3.50 & 2.84$\pm$2.96 & 1.34$\pm$0.85 & 0.47$\pm$0.33 & 3.35$\pm$2.78 & \ldots\FBb & 0.57\\ [3pt]
    24 & $8.16^{+0.09}_{-0.12}$ & 4.08$\pm$1.77 & 3.79$\pm$1.44 & 2.26$\pm$0.76 & 0.76$\pm$0.51 & 0.28$\pm$0.19 & 3.16$\pm$2.77 & 104$\pm$62 & 1.02\\ [3pt]
    25 & $8.73^{+0.19}_{-0.33}$ & 4.12$\pm$6.10 & 3.83$\pm$3.67 & 2.95$\pm$3.02 & 1.59$\pm$0.94 & 0.57$\pm$0.37 & 3.41$\pm$2.80 & \ldots\FBb &0.55\\ [3pt]
    26 & $9.22^{+0.11}_{-0.14}$ & 158.32$\pm$156.20 & 81.72$\pm$76.19 & 81.05$\pm$76.76 & 2.74$\pm$2.08 & 1.10$\pm$0.88 & 3.06$\pm$2.76 & \ldots\FBb & 0.14\\ [3pt]
    27 & $8.43^{+0.13}_{-0.18}$ & 9.08$\pm$10.38 & 7.00$\pm$5.51 & 5.06$\pm$4.98 & 2.04$\pm$0.92 & 0.78$\pm$0.40 & 3.17$\pm$2.78 & \ldots\FBb & 0.11\\ [3pt]
    28 & $8.57^{+0.12}_{-0.16}$ & 8.37$\pm$4.55 & 7.79$\pm$3.50 & 4.93$\pm$2.03 & 0.96$\pm$0.44 & 0.35$\pm$0.17 & 3.22$\pm$2.78 & 137$\pm$108 & 0.32\\ [3pt]
    \hline
    \end{tabular}
    \begin{tablenotes}
    \item \leavevmode\kern-\scriptspace\kern-\labelsep (1): DEEP2 \OIII 4363 galaxy ID. (2): Stellar mass. (3)--(5): SFRs measured instantaneously, averaged over the last 10 Myr, and averaged over the last 100 Myr, respectively. (6)--(7): Dust attenuation in the FUV- and V-bands, respectively. (8): e-folding time for star formation. (9): Age of galaxy. (10): Reduced $\chi^2$ for best fit. Uncertainties are reported at the 16th and 84th percentile.
    \item[a] \label{tn:2a} Ages are not well constrained.
    \item[b] \label{tn:2b} SED model not well constrained due to the limited photometric dataset (e.g., 3 bands, $BRI$, and 1 limit, [3.6]).
    
    \end{tablenotes}
  \end{threeparttable}
\end{table*}

\begin{table}
  \centering
  \caption{Binned DEEP2+\MACT\ \MZ\ Relation}
  \label{tab:bin}
  \begin{threeparttable}
    \begin{tabular}{crcc}
    \hline\hline
    Bin & $N$ & $\left<\log(M_{\star}/M_{\odot})\right>$ & $\left<12+\logOH\right>$\\
    (1) & (2) & (3) & (4) \\\hline
      7.5$\pm$0.25 & 5 & $7.46^{+0.002}_{-0.01}$ & $7.63^{+0.20}_{-0.05}$ \\[0.1cm]
      8.0$\pm$0.25 & 19 & $7.99^{+0.04}_{-0.03}$ & $7.92^{+0.08}_{-0.07}$ \\[0.1cm]
      8.5$\pm$0.25 & 36 & $8.52^{+0.02}_{-0.03}$ & $8.01^{+0.05}_{-0.05}$ \\[0.1cm]
      9.0$\pm$0.25 & 32 & $8.94^{+0.02}_{-0.03}$ & $8.19^{+0.05}_{-0.06}$ \\[0.1cm]
      9.5$\pm$0.25 & 9 & $9.42^{+0.03}_{-0.04}$ & $8.38^{+0.13}_{-0.10}$ \\[0.1cm]
      10.0$\pm$0.25 & 3 & $9.92^{+0.04}_{-0.11}$ & $8.37^{+0.23}_{-0.20}$ \\\hline
    \end{tabular}
    \begin{tablenotes}
    \item \leavevmode\kern-\scriptspace\kern-\labelsep (1): Stellar mass bin. (2): Number of galaxies. (3): Average $\log(M_{\star}/$\Msun$)$. (4): Average 12 + $\logOH$. Uncertainties are reported at the 16th and 84th percentile, and are determined using the bootstrapping method (described in Section \ref{sec:relation}).
    \end{tablenotes}
  \end{threeparttable}
\end{table}

\newcommand{\FAb}{\tnotex{tn:1b}}
\begin{table}
  \centering
  \caption{Best Fit for the $z\sim1$ \MZ\ Relation}
  \label{tab:best_fit}
  \begin{threeparttable}
    \begin{tabular}{ccc}
      \hline\hline
      $\log\left(\MTO/M_{\odot}\right)$ & 12 + $\logOH_{\rm asm}$ & $\gamma$\\
      (1) & (2) & (3)\\\hline
      $8.74^{+0.43}_{-0.83}$ & $8.45^{+0.15}_{-0.25}$ & $0.59^{+0.15}_{-0.21}$ \\[0.1cm]
      8.901\FAb            & $8.51^{+0.06}_{-0.05}$ & $0.59^{+0.09}_{-0.09}$ \\\hline
    \end{tabular}
    \begin{tablenotes}
      \item \leavevmode\kern-\scriptspace\kern-\labelsep Fitting parameters of the \MZ\ relation: (1) turnover mass, (2) asymptotic metallicity at the high stellar mass end, and (3) slope at the low-mass end. Uncertainties are reported at the 16th and 84th percentile, and are determined from the probability functions marginalized over the other two fitting parameters.
      \item[a] \label{tn:1b} $\MTO$ was held fixed to the value determined by \citetalias{AM13}.
    \end{tablenotes}
  \end{threeparttable}
\end{table}

\newpage

\begin{appendix}
\section{Mass--Metallicity--SFR Plots}

\begin{figure*}
	\includegraphics[width=7.0in,keepaspectratio]{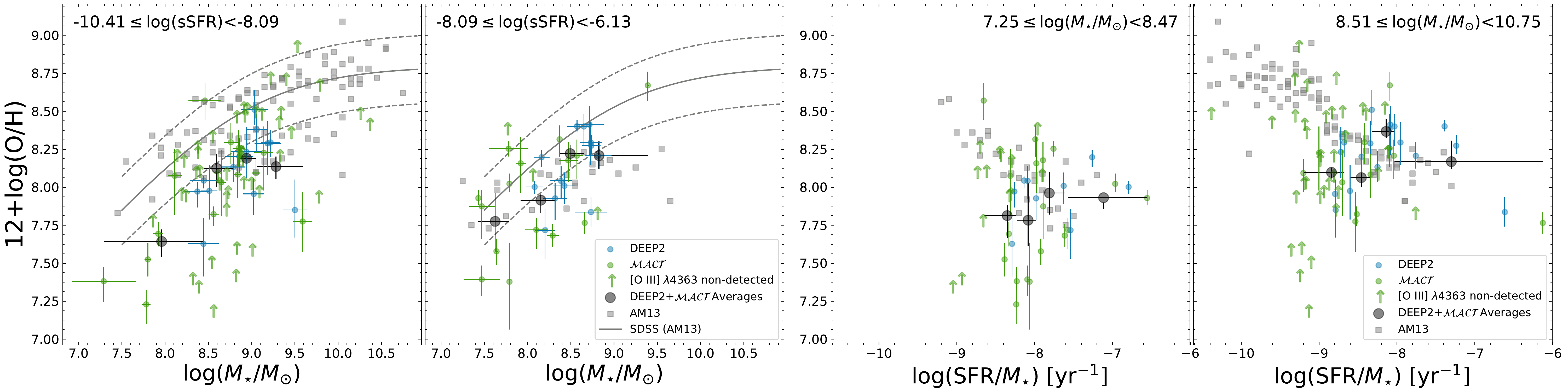}
    \vspace{-0.5cm}
    \caption{Oxygen abundance as a function of stellar mass (sSFR $\equiv$ SFR/$M_{\star}$) binned along sSFR (stellar mass) illustrated in the left (right) panel. Blue circles are \z\ DEEP2 galaxies from this study. Green circles and green arrows are the \MACT\ \OIII 4363-detected and non-detected samples, respectively. Black circles are the average metallicity and bootstrapped errors in bins with equal number (6--7) of \OIII 4363-detected galaxies. The range of the $x$ error bars is based on the minimum and maximum mass (sSFR) in each bin. Galaxies from \citetalias{AM13} are overlaid as grey squares. In the left panel, the \MZ\ relation from \protect\citetalias{AM13} is shown by a solid grey line, with grey dashed lines enclosing $\pm$1$\sigma$. These projections illustrate that a strong dependence of the \MZ\ relation on sSFR, as seen in local galaxies, is not seen for these \z\ galaxies.}
    \label{fig:bin_mass_SFR}
\end{figure*}

\begin{figure*}
	\includegraphics[width=\columnwidth,keepaspectratio]{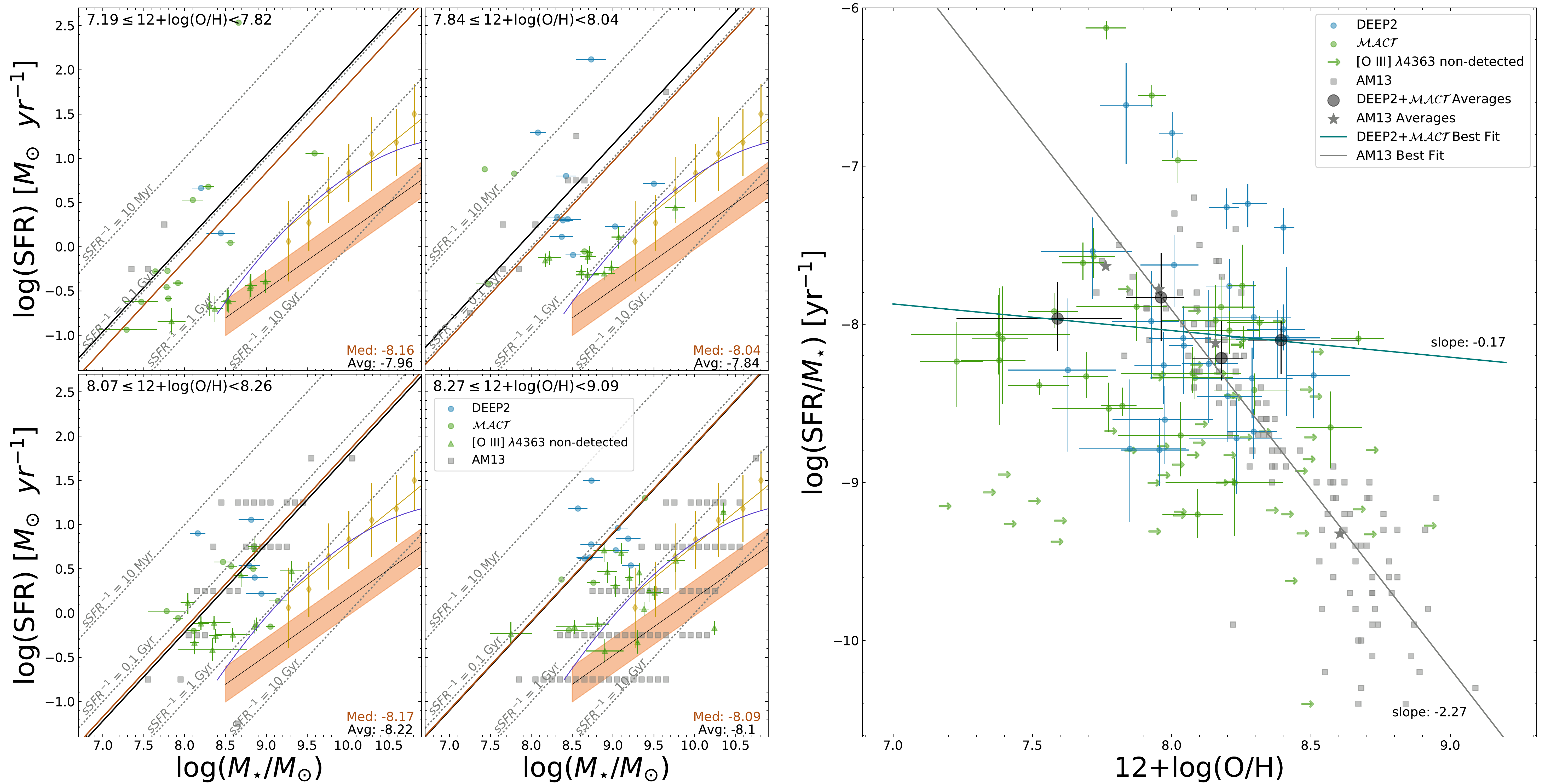}
    \vspace{-0.5cm}
    \caption{Left: SFR as a function of stellar mass for sub-samples binned according to oxygen abundance. Colour and point style schemes for galaxies follow those in Figure \ref{fig:bin_mass_SFR}. The $M_{\star}$--$\rm{SFR}$ relation, determined by \protect\cite{Salim07} at $z\sim0.1$, \protect\cite{Reyes15} at $z\approx0.8$, and \protect\cite{Whitaker14} at $0.5<z<1$, are plotted as an orange band, yellow diamonds, and a purple line, respectively. Lines of constant inverse specific SFR (sSFR$^{-1}$) are shown as dashed grey lines with corresponding timescales. The median (average) sSFR$^{-1}$ of the DEEP2+\MACT\ galaxies is shown as a brown (black) line, and the average is illustrated in the right panel. Right: Specific SFR as a function of metallicity. The black circles are the average sSFR in bins with equal number (13--14) of \OIII 4363-detected galaxies. Here the sSFR error bars are the bootstrapped averages while the metallicity error bars are based on the minimum and maximum in those bins. The turquoise (grey) line illustrates the least-squares fit for DEEP2+\MACT\ (\citetalias{AM13}). As shown, sSFR increases as metallicity decreases, but at a shallower slope than local galaxies. For DEEP2+\MACT, we performed t-tests that demonstrated that we cannot rule out zero sSFR--$Z$ dependence at $>$95\%, but we are able to rule out the steep dependence found for local galaxies at $>$4$\sigma$ significance.}
    \label{fig:bin_met}
\end{figure*}

In this appendix, we provide different projections of the mass--metallicity--SFR relation. Figure \ref{fig:bin_mass_SFR} compares gas metallicity against stellar mass (left panel) and sSFR (right panel). Here, binning was performed along sSFR and stellar mass, respectively.  In Figure \ref{fig:bin_met}, we illustrate SFR as a function of stellar mass in different metallicity bins (left panel), and how the sSFR depends on metallicity (right panel).

\end{appendix}

\bsp	
\label{lastpage}

\end{document}